\documentclass[iop]{emulateapj}
\usepackage{float}
\usepackage{natbib}
\usepackage{epsfig}
\usepackage{graphicx}
\usepackage{hyperref}

\shorttitle{ WASP-14b Snapshots}
\shortauthors{Krick et al.}

\begin{document}
\newcommand\msun{\hbox{M$_{\odot}$}}
\newcommand\lsun{\hbox{L$_{\odot}$}}
\newcommand\magarc{mag arcsec$^{-2}$}

\bibliographystyle{apj}
\title{\bf Spitzer IRAC Sparsely Sampled Phase Curve of the Exoplanet WASP-14b}

\author{J.E.~Krick\altaffilmark{1}, J.~Ingalls\altaffilmark{1},
  S.~Carey\altaffilmark{1}, K.~von Braun\altaffilmark{2} ,
  S.R.~Kane\altaffilmark{3}, D.~Ciardi\altaffilmark{1},
  P.~Plavchan\altaffilmark{4}, I.~Wong\altaffilmark{5}, P.~Lowrance\altaffilmark{1}}

\altaffiltext{1}{Spitzer Science Center, MS 314--6,
California Institute of Technology, Jet Propulsion Laboratory,
Pasadena, CA 91125, USA}
\email{jkrick@caltech.edu}
\altaffiltext{2}{Lowell Observatory
Flagstaff, AZ, USA}
\altaffiltext{3}{San Francisco State University
San Francisco, CA, USA}
\altaffiltext{4}{Missouri State University
Springfield, MO, USA}
\altaffiltext{5}{California Institute of Technology,
Pasadena, CA, USA}

\begin{abstract}

  Motivated by a high Spitzer IRAC oversubscription rate, we present a
  new technique of randomly and sparsely sampling phase curves of hot
  Jupiters.  Snapshot phase curves are enabled by technical advances
  in precision pointing as well as careful characterization of a
  portion of the central pixel on the array.  This method allows for
  observations which are a factor of roughly two more efficient than
  full phase curve observations, and are furthermore easier to insert
  into the Spitzer observing schedule.  We present our pilot study
  from this program using the exoplanet WASP-14b.  Data of this system
  were taken both as a sparsely sampled phase curve as well as a
  staring mode phase curve.  Both datasets as well as snapshot style
  observations of a calibration star are used to validate this
  technique.  By fitting our WASP-14b phase snapshot dataset, we successfully
  recover physical parameters for the transit and eclipse depths as
  well as amplitude and maximum and minimum of the phase curve shape
  of this slightly eccentric hot Jupiter.  We place a limit on the
  potential phase to phase variation of these parameters since our
  data are taken over many phases over the course of a year.  We see
  no evidence for eclipse depth variations compared to other published
  WASP-14b eclipse depths over a 3.5 year baseline.

\end{abstract}


\section{Introduction} 
\label{intro}


The study and characterization of exoplanetary atmospheres has been,
and continues to be, one of the greatest legacies of the {\it Spitzer} Space
Telescope \citep{2004ApJS..154...10F,2004ApJS..154....1W}.  The
unique capabilities of {\it Spitzer}, coupled with the favorable planet-to-star
contrast ratio in the infrared \citep{2006ApJ...650.1140B}, has
allowed numerous achievements to date.  For the first time, photons
from a planet outside our solar system were directly
detected through observations of a planet's secondary eclipse which gave us the
first insights into the temperatures of the planet's synchronously
rotating dayside
\citep[e.g.,][]{2005ApJ...626..523C,2005Natur.434..740D}. Emission
spectrophotometry produced the first evidence of water molecules in the
dense exoplanetary atmospheres
\citep[e.g.,][]{2008Natur.456..767G}. The Infrared Array Camera (IRAC)
high precision photometry produced pioneering characterization of the
atmospheres of extrasolar planets \citep[e.g.,][]{2007Natur.447..183K}.  While
observations of secondary eclipses have been performed using
ground-based facilities
\citep[e.g.,][]{2009A&A...493L..31S,2012ApJ...744..122Z}, the study of
phase curves and the dynamics of exoplanetary atmospheres has only been
achieved from space.

Coupled with theoretical modeling
\citep[e.g.,][]{2005ApJ...629L..45C,bur06,2008ApJ...678.1436B,cow08,2010ApJ...720..344L,2010ApJ...719..341B,2010ApJ...713.1174M,2010ApJ...716..144T,cow11} and {\it Spitzer}
observations of secondary eclipses, phase curves can be used to put
constraints on various parameters of the exoplanet atmosphere such as
the atmospheric pressure structure, chemical composition, and the
combination of albedo, different opacities, and existence of a
temperature inversion in the upper layers of the atmosphere. 

A significant fraction of extrasolar planets whose atmospheres have been characterized to
date are 'hot Jupiters', i.e., planets approximately the size and mass
of Jupiter that orbit their respective parent star once every few
days. Their rotation rate is consequently believed to be synchronized with
their orbital period, such that the same side of the planet always faces the
star. These exoplanets receive approximately 10,000 to 100,000 times the flux
received by Jupiter which means their dayside temperatures reach
values as high as several thousand Kelvin, while the nightside
temperatures can be hundreds to thousands of Kelvin cooler.
Additionally, hot Jupiters represent a simpler picture for
circulation models in that they are in chemical equilibrium, largely
cloud free, and there is not expected to be convection in their
atmospheres \citep{2006ApJ...652..746F}.  Understanding a sample
of hot Jupiters is a vital stepping stone on the path to future
characterization of cooler planets, including those in the habitable
zones, by future NASA space missions such as JWST.

About a dozen infrared phase curves have been published to date (see review in \citet{2015ApJ...811..122W}\citep{2006Sci...314..623H,2010ApJ...723.1436C,2007Natur.447..183K,2009ApJ...690..822K,2012ApJ...754...22K,2009ApJ...703..769K,2009Natur.457..562L,2013ApJ...776L..25D,2012ApJ...747...82C,2012ApJ...760..140C,2013MNRAS.428.2645M,2013ApJ...766...95L,2007MNRAS.379..641C,2012ApJ...752...81C,2014ApJ...790...53Z,2015ApJ...811..122W,2015arXiv151209342W}. The next scientific
breakthrough in this field will come from exploring a large sample of
extrasolar planets and determining how their atmospheric properties
depend on physical conditions: comparative atmospheric sciences.  The
shape of the infrared phase curve can be interpreted as a longitudinal
brightness temperature distribution across the planet and provides the
best measurement of the efficiency of the energy transport in the
atmosphere. There are many factors involved in understanding the energy
redistribution from the day- to nightside
\citep{bur06,2008ApJ...678.1436B,cow11}. The measured energy emitted
by the planet should equal the energy received by the planet from the
star, modulo non-zero Bond albedo values and assuming no residual
energy from planet formation \citep{knu09b}. The longitudinal
temperature distribution is driven by three effects: (1) the time it
takes to absorb and reemit flux in the planet'€™s atmosphere (radiative
timescale), (2) the time it takes to transport a parcel of heated
gas to the planet's nightside (advective timescale), and (3) chemistry
which can effect the pressure level of the atmosphere probed at any wavelength. Any potential
shift in time between the brightness maximum of the phase curve and
the time of secondary eclipse (ie., when the substellar spot is
visible) is an indicator of the relative strength of radiative
(heating/cooling) and advective(wind) processes. In addition, there are deviations
in the planetary emission from a blackbody spectrum, such as inversion
layers at the observed (wavelength-dependent) height in the
atmosphere, which may affect the longitudinal temperature
distribution.

The interpretation of published phase curves \citep[e.g..][]{cow11,2013ApJ...776..134P,2015MNRAS.449.4192S,2015ApJ...811..122W} has clearly
illustrated a diversity in atmospheric circulation patterns within the
exoplanet population studied thus far. The reasons for this are not
clear, principally due to a lack of data. We hope to add data to this
effort, starting with this first paper in a series to study phase
curves of hot Jupiters through a novel snapshot technique which will
help us compile a large sample with less observing time invested on
the highly oversubscribed {\it Spitzer} Space
Telescope.

There are two new techniques presented in this paper, 1) using a gain
map dataset to remove intrapixel gain effect in the data and 2)
sparsely sampled phase curves.  The advent of precision pointing and
characterization of the intrapixel sensitivity of the central pixel in
the IRAC subarray make this sort of project possible.  Phase curve
observations do not have to be conducted in the time-consuming manner of
multiple-day, consecutive observations. Instead, given that we know
the periods of the planets, it is more efficient to build up phase
curves over time from many, randomly spaced, short duration
observations (see Section 2 for a more detailed description).  The
slew time of the {\it Spitzer} space telescope is relatively fast, so there
is no large penalty to observing targets with multiple epochs.

Something similar to our snapshot technique was used in
\cite{2007MNRAS.379..641C}.  However, those observations nodded back
and forth between the science target and a flux calibration star.
This prevented them from making a correction for the intrapixel gain
effect because the nods end up on different positions on the pixel.
We have developed novel approaches, described below, in both observation
planning and data reduction to correct for the intrapixel gain effect
and other instrumental noise sources.

We present our pilot study with these new techniques of the hot
Jupiter WASP-14b, a 7.3 Jupiter mass planet with  eccentricity of 0.083
\citep{2015ApJ...811..122W} orbiting an F5V star at a
radius of 0.036 AU with a 2.24 day period \citep{2013ApJ...779....5B,
  2011A&A...529A.136E}. The star has a brightness of $K$=8.6 with $T_{eff}
= 6480\pm140K$ and solar metallicity \citep{2009MNRAS.392.1532J}. The
above properties make this planet a prototype of hot
Jupiters. \cite{2009MNRAS.392.1532J} note a relatively high density
for this planet of 4.6 g cm$^{-3}$ given that its radius is $1.26
R_{Jupiter}$ radii.  WASP-14b is known to have a significant
spin-orbit misalignment, which, coupled with its eccentricity, is
indicative of the orbital evolution of this massive planet
\citep{2009PASP..121.1104J}.

WASP-14b is predicted to have a thermal
inversion based on its high level of irradiation and the activity of
its parent star
\citep{2008ApJ...678.1419F,2010ApJ...720.1569K}, however \cite{2013ApJ...779....5B} find
that eclipse spectroscopy at 3.6, 4.5, and $8.0\micron$ are well fit
with no thermal inversion. \cite{2012ApJ...758...36M} attribute the
shape of the broad-band eclipse spectrum to possible condensation and
gravitational settling of the TiO and VO \citep{2009ApJ...699.1487S}
or a carbon-rich atmosphere with naturally low TiO and
VO. \cite{2013ApJ...779....5B} find evidence for a relatively low
($<30\%$) day-night energy redistribution and that the dayside
spectrum of WASP-14b is consistent with both carbon-rich and
oxygen-rich chemistry, with the latter being a marginally better fit.

This paper is structured in the following manner.  Section 2 describes
our strategy for snapshot observations including a discussion of
stellar variability.  Section 3 discusses
systematic effects in IRAC data and our mitigation techniques.  Data
are described in Section 4 and data reduction methods are explained in
Section 5.  Results and Discussion are in Section 6 and the paper
concludes in Section 7.  Throughout the paper we will refer
interchangeably to 'Ch1' or $3.6\micron$ and 'Ch2' or $4.5\micron$.


\section{Snapshot Strategy}
\label{snaps}

``Snapshot Phase Curves'' is a program to observe a planet-hosting star for one
epoch, and then later re-point to that star and observe it for another
epoch, repeated at random intervals determined by brief holes in the
{\it Spitzer} observing schedule, until we have built up a well sampled phase
curve. Building phase curves in this way keeps data volumes low and
does not require the observatory to be taken over for days at a
time, as for traditional {\it Spitzer} exoplanet phase curve measurements.  Snapshots
are more efficient by a factor of two to three than long continuous staring mode
observations, depending on the period, and are easier to schedule,
which is especially important as the mission goes forward.

We set the duration of a single epoch snapshot at 30 minutes.  This
timescale is chosen for two reasons; (1) to stay within the roughly
30 - 40 minute pointing wobble (see \citet{IRACHPP}
for a discussion of spacecraft motions), and (2) it allows us to build up
enough signal when binning to reach our signal to noise ratio (SNR)
goals (described below).

We investigate the number of snapshot-style observations required to
adequately sample a phase curve.  We constructed a Monte- Carlo
simulation of phase curves and simulated datasets that assume a random
distribution of 30-min long snapshots (termed Astronomical Observation
Requests (AORs) by {\it Spitzer}) in orbital phase-space.  These
simulated datasets are very simple approximations of IRAC photometry
with a noise level commensurate with expected IRAC noise for a
specific brightness star (this simulation was made prior to the more
detailed IRAC
data simulator\citep{2012SPIE.8442E..1YI}).  The success of a
simulation is judged by how well the phase amplitude measured on the
simulated dataset matches the real (input) amplitude. Assumptions
which go into this include (1) an expected phase curve amplitude of
0.06\% (the smallest amplitude predicted for known hot Jupiters at the
time of observation planning in 2011), (2) a photometric uncertainty per
datapoint of 0.02\% in relative flux, (3) a sinusoidal shape of the
phase curve \citep{cow07}, and (4) that the combined data for a given
target cover a time-span much longer than the period of the
exoplanet. Assumption (4), combined with a short AOR duration, ensures
that the simulation results are independent of the planetary
orbital duration. Assumption (2) of photometric uncertainty is based on
signal-to-noise calculations for 30-min on a 25 mJy source, which is a
factor of three fainter than WASP-14b to accommodate all possible
targets in our extended program.

\begin{figure}
\includegraphics[angle=270, scale=0.3]{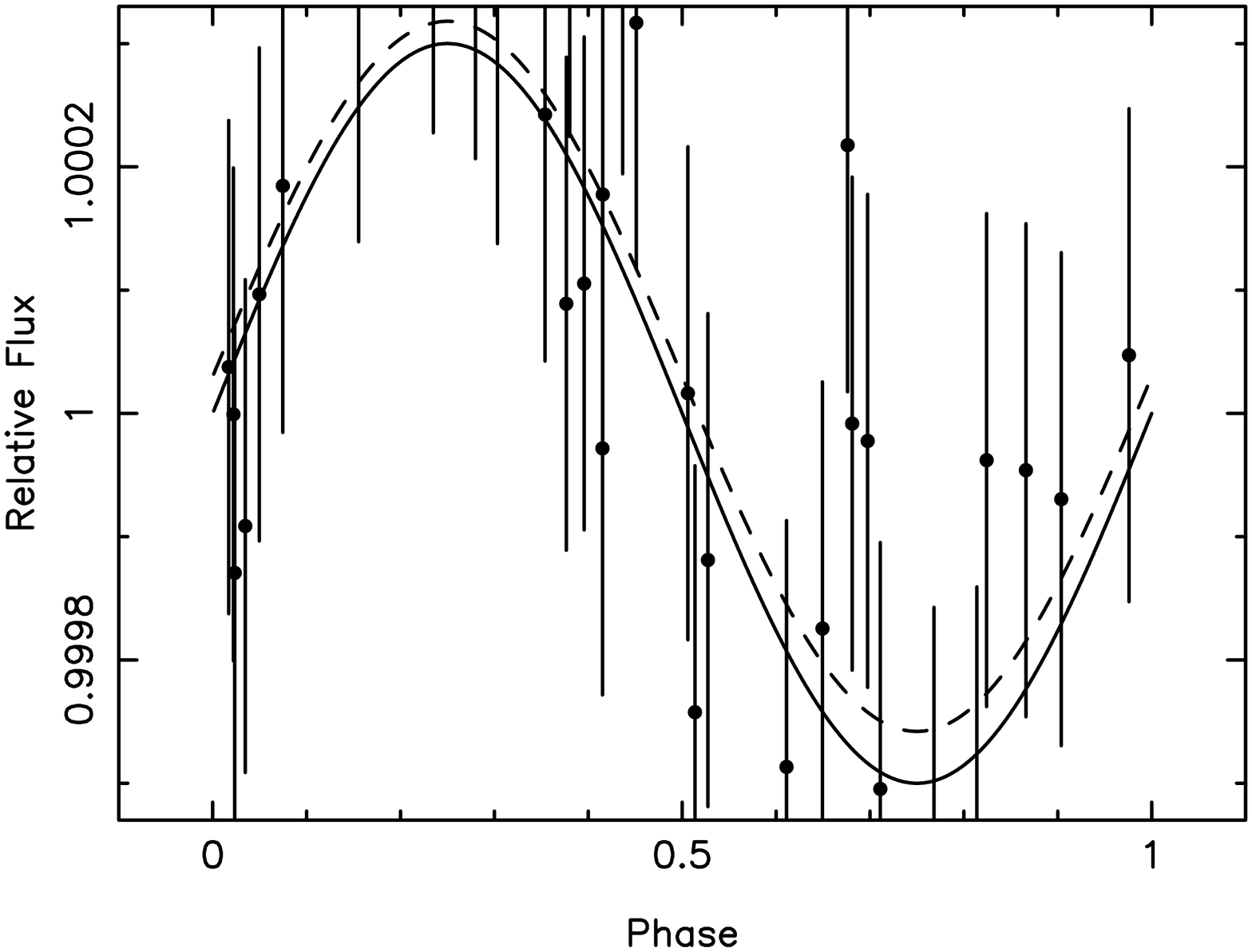}  
\includegraphics[angle=270, scale=0.3]{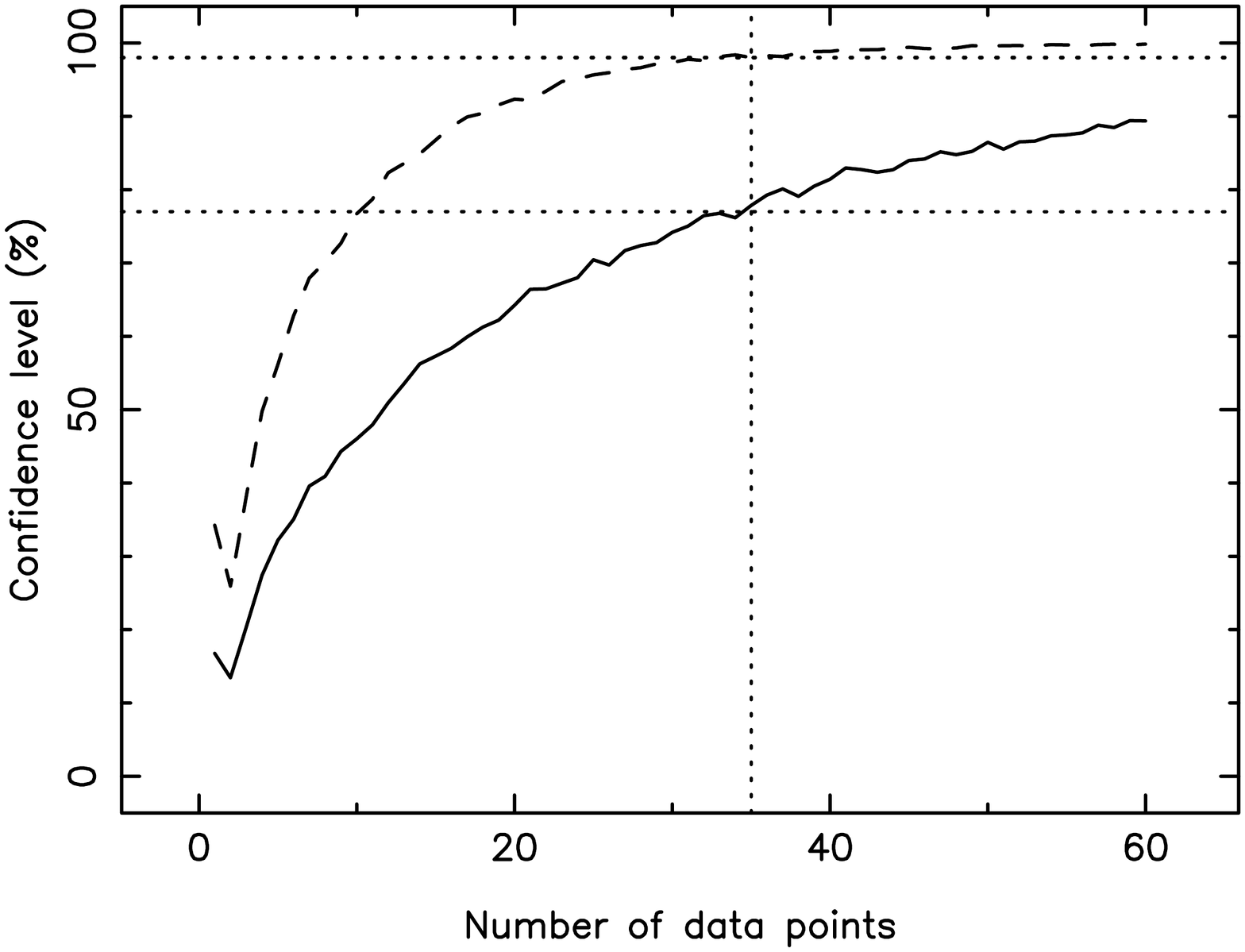}  
\caption[sim_number]{Number of Required Observations: Left panel:
  Example of simulation: shown are the 'true'€ phase curve (solid
  line), 35 randomly simulated, phased data points (30-min long per
  point) with estimated error bars (based on a source which is one
  third as bright as our target), and the corresponding fit to these
  data points (dashed line) where the fitted amplitude differs from
  the input by 20\%. Right panel: Results of the simulations: number
  of data points necessary to recover the amplitude of the input phase
  curve to within 20\% (solid line) and 40\% (dashed line). The small
  tick at the beginning of the curves is due to the fact that the
  $\chi^2$ fit is unreliable for low data numbers.  This simulation
  finds that 35 epochs of observations per target can achieve 20\%
  (40\%) precision in characterizing the phase curve at the 77\%
  (99\%) confidence level for these faintest targets, and will do much
  better for brighter ones such as WASP-14b.  See \S \ref{snaps} for
  more details. }
\label{fig:sim_number}
\epsscale{1}
\end{figure}

In our simulation, the number of snapshot-style AORs was incremented
from 1 to 60. The top range of the number of simulated AORs represents
a rough boundary where it might not be worth using this technique
because the overall time required for snapshots would be similar to a
continuous phase curve (depending on the period of the planet).  We
generated the Monte-Carlo simulation with 5000 datasets, or sets of
AORs, for each number of snapshot-style AORs bin.  A $\chi^2$ fit was
performed for each dataset and the result was compared to the input
parameters. We show the results in Figure \ref{fig:sim_number}. The
left panel illustrates an example simulated dataset of 35
measurements, where the solid line is the model used to generate the
data and the dashed line is the best-fit to these data.  The right
panel summarizes all of the simulations. The dashed line is for those
simulations where $\chi^2$ fits produced an amplitude within 40\% of
the actual amplitude, constituting a marginal detection, but
nevertheless providing a quantitative upper limit on exoplanetary
day/night contrast. The solid line indicates the percentage of
datasets for which a measured amplitude within 20\% of the actual
amplitude was recovered. At this level of precision, much more
quantitative constraints can be imposed on atmospheric
properties. Thus, for 35 measurements per target, we can recover the
amplitude of the phase curve to within 20\% of its input value 77\% of
the time and to within 40\% of the input value 99\% of the time.

After observing 35 WASP-14b AORs in September 2012, we noticed that
the assumed randomness of the observations was not achieved, and thus
scheduled another 11 AORs in the hopes of filling in phase regions
which were not covered in the original 35 AORs.  46 total snapshots
provide a well-sampled phase curve with fortuitous observations
of one AOR during transit and two AORs during eclipse.

\subsection{Stellar Variability}
\label{stellar_var}
If the star varies, then a snapshot light curve might just be
capturing stellar variation instead of planet phase variations. The
planetary variations are temporally well separated from stellar
variations because the timescales of planetary orbits or rarely
similar to the timescales of stellar rotations.  The concern is that
because we observe over many months and many phases, we may see an
offset in fluxes from one snapshot to the next due to stellar
variability.  We chose WASP-14b because statistically it's spectral
type (F5V) indicates it should be relatively quiescent based on
average variability of stars in the Kepler input catalog, the
periodogram of the original survey data, and the measured radial
velocity (RV) residuals.

Looking at a sample of F stars in
the Kepler Input Catalog, \citet{2011AJ....141..108C} find an average
dispersion in 30 minute bins over 33 days of 0.1~mmag in the optical for F
stars with Kepler mags in the range of WASP-14b ($K_{p}$ estimated to be 9.7).  We
expect the IR variation to be smaller than these average optical
variations.  Also, based on an emission spectrum,
\citet{2010ApJ...720.1569K} find that WASP-14 is inactive.

\cite{2013ApJ...779....5B} examine the periodogram of the original
WASP light curve data to search for periodic signals as evidence of stellar
activity.  They find no significant (false alarm probability of less
than 0.05) periodic signals in that dataset, implying no measurable
stellar variability.

Lastly, stellar activity in the form of non-radial pulsations, or
inhomogeneous convection or spots can cause RV variations which might
show up in RV surveys published to date \citep{2000A&A...361..265S}.
\cite{2009MNRAS.392.1532J}, in the original discovery paper, quote RV
residuals of 10.1m/s .  \citet{2015ApJ...811..122W} find a residual of
12.3 m/s based partially on data from \citet{2014ApJ...785..126K}.
This is in the expected range for this type of star, and is not high
enough to suggest stellar activity ($>40m s^{-1}$)
\citep{2004AJ....127.3579P}.

\section{IRAC Sources of Noise}
\label{noise}

\begin{figure}
\hspace*{-2.5cm}
\vspace*{-5.5cm}
\includegraphics[scale=0.7]{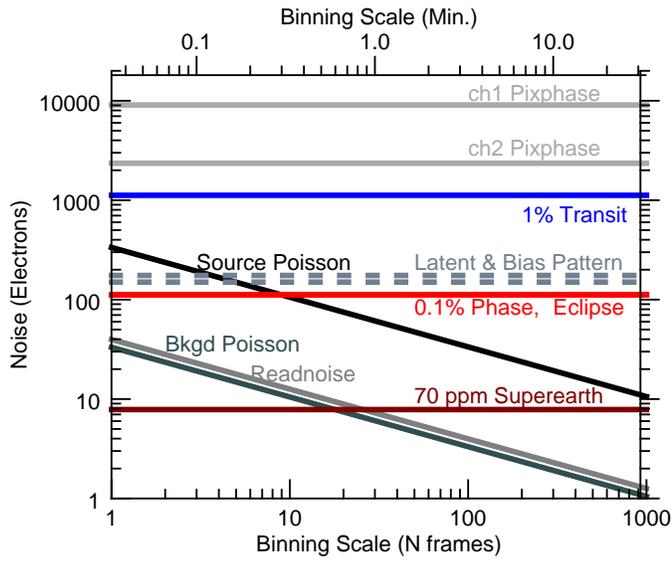}  
\caption[noise_sources]{Noise sources in IRAC high precision
  photometry.  Black diagonal line shows the poisson noise due to a
  half-full well source.  Noise due to that source's background and
  readnoise are shown in gray diagonal and are negligible for this
  work.  The strength of the intrapixel gain effect is shown for both
  channels as horizontal light gray lines.  Note that we choose to
  observe in Ch2 in large part because its intrapixel gain
  effect is lower than Ch1.  Horizontal lines for 1\%, 0.1\%,
  and 70~ppm effects are shown for reference.  Also noted as dashed
  lines are estimates for where latent and bias pattern noise will
  effect the photometry. See \S \ref{noise} for more details.}
\label{fig:noise_sources}
\epsscale{1}
\end{figure}

Data reduction for high precision photometry is challenging due to
both instrumental and astrophysical effects.  Instrumental effects are
discussed below and their relative strengths are shown in Figure
\ref{fig:noise_sources}.  This figure includes the Poisson
contribution from an assumed source at half-full well and the
background along with the readnoise. The intrapixel gain effect for
each channel is shown as a grey line and discussed in section
\S\ref{pixphaseeffect}.  Because this is such a strong effect, it must
be removed from the data before we can detect phase
variations.

Estimates of the strength of the effect on photometry of latent images
and a detector bias pattern are shown as dashed gray lines.  Low level persistent images exist
after a bright star has been observed, and can last for many hours.
These are discussed in \S \ref{latent}.  Changing bias patterns
can be generated by a difference in the delay times before images are
recorded from the darks to the science frames.  The bias pattern
effect has not been fully characterized and cannot be derived
from this dataset.  The estimates of the bias effect used in this
plot are determined from a set of five 5 - 10 hour continuous staring mode
archival observations of a blank field (no star targeted).  The level
of this bias effect is uncertain and will vary as a function of time
over the days to months between observations presented here.

As examples of the rough level of the signals relevant to exoplanet
studies with IRAC, we also show the relative levels of a 1\% transit depth,
a 0.1\% eclipse depth, and a 70 ppm postulated Super-Earth secondary
eclipse depth.

\subsection{Intrapixel Sensitivity }
\label{pixphaseeffect}
\label{pmap}

Due to the under-sampled nature of the PSF, the warm IRAC arrays show
variations of as much as $8\%$ in sensitivity as the center of the PSF
moves across a pixel due to normal spacecraft motions
\citep{2012SPIE.8442E..1YI}. These intra-pixel gain variations are the
largest source of correlated noise in IRAC photometry (see figure
\ref{fig:noise_sources}). Many data reduction techniques rely on the
science data themselves to remove the gain variations as a function of
position\citep{cha05,2010PASP..122.1341B,2012ApJ...754..136S,2012ApJ...754...22K,2013ApJ...766...95L,2015ApJ...805..132D}.
The limitation of self-calibration techniques is that it does not work
well for sparsely sampled dataset (many full phase curves).  The SSC
has generated a high-resolution gain map from standard star data
\citep{2012SPIE.8442E..1YI}, which can be used as an alternative when
reducing data.

The use of the gain map reduction technique is reliant on PCRS
Peak-Up.  PCRS Peak-Up uses the {\it Spitzer} Pointing Calibration Reference
Sensor (PCRS) to repeatedly position a target to within 0.25 IRAC
pixels of an area of minimal gain variation.  It is important to land
on this 'sweet spot'€ ([15.120,15.085] in ch2) because it a) minimizes the effect on the photometry of
standard telescope motions, b) it is the most well-calibrated position
on the detector, and c) it enhances measurement repeatability from one epoch to the next.  The
minimization of the intra-pixel gain effect happens both because that
region is well calibrated, and because the slope of the gain map is
shallowest at this position.  The change in gain as a function
of position is minimized to $\sim0.2\%$ across the sweet spot, which
is smaller than at other positions on the pixel.

Most staring mode exoplanet observations employ the SSC recommended
practice of observing a 30 min. pre-AOR to allow the telescope to
settle.  In the interest of time, we do not observe this pre-AOR, and
instead peak-up directly to our target and start the snapshot
observations.  

The gain map  (``pmap'') dataset uses the IRAC calibration stars KF09T1 in ch1 and
BD+67 1044 in ch2 taken with subarray 0.4 and 0.1s frame times
respectively.  The Feb 26 2015 version of the $4.5\micron$ pmap dataset
used in this reduction includes a total of 409,539 photometry points,
90\% of which are within the sweet spot.  Initial mapping of the
central pixel deliberately included the whole pixel as work was beginning to define
the sweet spot.  In general PCRS Peak-Up places a target within the
sweet spot 98\% of the time \citep{2012SPIE.8442E..1YI}.  This paper
only uses ch2.  All gain map data are reduced in the same manner as
the science observations.

The limitation of using a gain map for removing gain variations from the
data is that it cannot be applied to data with positions off the sweet
spot.  This is more often the case for targets which are faint or
extremely bright, and high proper motion stars.  We leave a
comparison of the various data reduction and analysis
techniques for a future study.
 
\subsection{Photometric Stability}
\label{degrade}
The snapshot technique only works if aperture photometry is stable as
a function of time, on year-long timescales.  This is because both the
gain map calibration dataset and the observations of WASP-14b were
taken over many years.  The gain map calibration dataset (see
\S\ref{pmap})was observed starting in 2011, with additional data taken
as recently as winter 2014 and the data on WASP-14b was taken over a
1.5 year baseline (see \S\ref{observations}.)  The calibration dataset
needs to be both consistent over the years it was observed and capable of
correcting data taken at any time during the mission.

\begin{figure}
\hspace*{-2cm}
\vspace*{-4.5cm}
\includegraphics[scale=0.7]{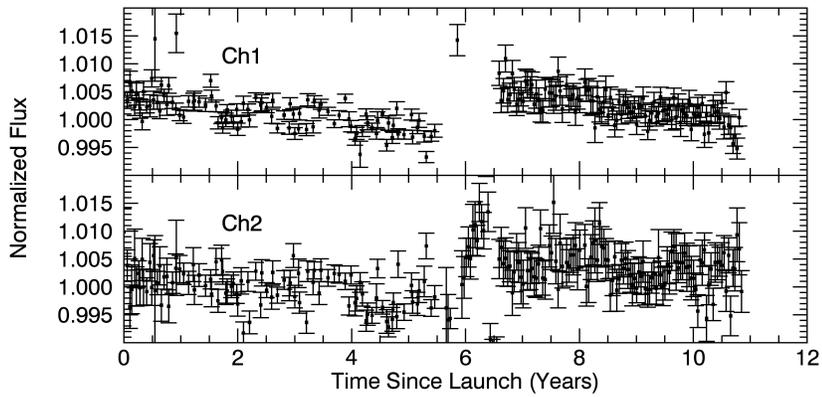}  
\caption[flux_stability]{ Photometric Stability as a Function of Time:
$3.6\micron$ on the top, and $4.5\micron$ on the bottom.  This plot shows
aperture photometry of the ensemble of calibration stars over the
entire mission to date.  There is a slight degradation as a function
of time which is possibly due to radiation damage to the optics.  The
gap around six years is the transition from cryo to warm where for a
short time we used a different calibration which will effect the
aperture fluxes.  See \S \ref{degrade} for more details.}
\label{fig:flux_stability}
\epsscale{1}
\end{figure}

To understand the photometric stability of the IRAC detectors as a
function of time, we examined the existing calibration data taken over
the course of the entire 11 year mission to date. Figure
\ref{fig:flux_stability} shows aperture photometry of seven IRAC
primary calibration stars binned together on two week timescales.
Primary calibrators are described in detail in
\cite{2005PASP..117..978R}.  Their main function is to determine the
absolute calibration of the instrument.

There is a statistically significant decrease in sensitivity of both
ch1 and ch2 over the course of the mission.  The degradation is
extremely small, of order 0.1\% per year in ch1 and 0.05\% per year in
ch2.  The decrease in sensitivity is potentially caused by radiation
damage to the optics.  Individual light curves for each of the
calibration stars used in this analysis were checked to rule out the
hypothesis that one or two of the stars varied in a way as to be the
sole cause of the measured decrease.  While the slope for each
individual star is not as well measured as for the ensemble of stars,
it is apparent that the decreasing trend is not caused by outliers.
We also rule out the solar cycle as the cause of flux degradation by
examining the cosmic ray rate as a function of time throughout the
mission.

In light of this very small flux degradation, we have corrected the
photometry of both the gain map dataset and the exoplanet host stars
by a linear function in time which decreases at 0.05\% per year.

\subsection{Persistent Images}
\label{latent}
Both IRAC channels sometimes show residual images of a source after it has been
moved off a pixel. When a pixel is illuminated, a small fraction of
the photoelectrons are trapped. The traps have characteristic decay
rates, and can release a hole or electron that accumulates on the
integrating node long after the illumination has ceased.  Persistent
images on the IRAC array start out as positive flux remaining after a
bright source has been observed. At some time later the residual
images turn from positive to negative, so that they are actually below
the background level (trapping a hole instead of trapping an
electron). Positive and negative persistent images in either the aperture or
background annulus in either the dark frame or the science frames can lead to
artificial increases or decreases in aperture photometry fluxes.  We
will use the terms persistent images and latents interchangeably
throughout the discussion.

We have learned from this work that persistent images are pervasive.
Short term persistent images start at about 1\% of the source flux and
can be seen to decay over the course of minutes to hours.  We look for
long term latents by median combining all warm mission darks at the 2s
frame time into a ``superdark''.  We also make yearly and seasonal
superdarks which are a median combine of a single year's worth of darks
or a single seasons (e.g.. Jan - Mar) worth of darks throughout the
mission.  Differencing the mission long superdark from the yearly and
seasonal darks reveals low-level long term latent patterns which
change from year to year and season to season. These residual latent
images are either caused by observing cadence (coincidentally {\it
  Spitzer} uses more subarray in the fall season) or by
long-lasting, low-level, persistent images.  Beyond knowing of their
existence, it is very difficult to track these low level latent images post
facto, especially since we know that the latents are both growing
with new observations dependent on the specific observing history, and
shrinking over time as they dissipate.

Long term latents are significant for this work because they change on
timescales over which our observations are made.  The effect on
snapshot photometry will have a random component, as the scheduling of
the snapshots was random. Our superdark analysis shows that persistent
images can effect the photometry of snapshots at the 0.1\% level (see
Figure \ref{fig:noise_sources}. In order to remove this effect for
future snapshot phase curve observations, we recommend observing a
dither pattern on a blank region before and/or after each
snapshot. This will ultimately provide a dark observation which is
specific for each snapshot observation.  The choice of observing the
dither before and after will depend on weather or not the persistent
image is stable on 30 min. timescales, of which we don't have a good
understanding. Adding these dithered observations will increase the time
required for these types of sparsely sampled phase curves. If the
persistent image level were changing significantly on the 30
min. timescale of the snapshots, we might expect to see its signature
in the RMS vs. Binning plots, which instead show only Poisson noise,
see \S\ref{rms}

This latent in the darks is not seen in the instrument stability
section (\S\ref{degrade}) where tests were done on photometry of calibration stars
because those observations are dithered around in position, as opposed
to holding the observatory in one single position.

\section{{\it Spitzer} Warm IRAC Observations }
\label{observations}

We present here snapshot observations of a pilot set of observations
of WASP-14b, which can be extended to more exoplanetary systems.  In
addition, we also discuss archival continuous staring mode observations of
WASP-14 and snapshot observations of the calibration star HD~158460
for verification of the observing strategy. Observing parameters are
listed in Table \ref{tab:obspars}.  All observations discussed in this
paper were taken with Warm IRAC in staring mode at $4.5\micron$ with
the target observed as close as possible to the sweet spot of the
central pixel of the sub-array.

We choose Ch2, and not Ch1, for both scientific and
technical reasons. First, the predicted planet-to-star flux ratio
values are larger at $4.5\micron$ (by tens of percent), which
increases the planetary contribution to the photometry. Second,
Ch2 is overall a better behaved instrument in the warm mission
(see Figure \ref{fig:noise_sources}). In particular, the pixel phase
effect is a smaller effect in Ch2 than Ch1
\citep{2012SPIE.8442E..1YI}. The only minor disadvantage to choosing
Ch2 that we are aware of is that the stars are fainter.


In Staring mode, no dithering or mapping is used; the
telescope is not intentionally moved after it arrives on target.  The
sub-array is a 32x32 pixel portion of the full detector array.  Images
are stored in sets of 64 sub-frames, tied together into one FITS file.
Data described in this paper are from all archival data on Program IDs
(PIDs) 80016, 80073, and calibration PIDs 1320, 1326, 1328,1331, 1333,
1336, 1338, 1346, 1658, 1659, 1669.

PCRS peak-up on the target was used prior to all AORs to reliably put
the target star onto the sweet spot, which is necessary for our
reduction technique.  We have only mapped the gain variations to
sufficient precision for this work over the single sweet spot of the
central subarray pixel in both channels (see \S \ref{pmap}).  If the target does not land
on the sweet spot where the existing gain map dataset can provide
calibration of the intrapixel gain effect, then this technique is not
possible, and snapshots can not be used to build phase curves.

For WASP-14b, a two second frame time was chosen from among the fixed
set available to observe this 68mJy star at $\sim38\%$ of full well,
where the IRAC detectors have the least non-linearity (see \S
\ref{nonlin}). In total, 46 snapshot visits were observed
randomly throughout two visibility windows separated by one year.  The
first 35 observations were taken from 2012-09-07 to 2012-09-27 UTC.
A further 11 snapshots were observed from 2013-08-24 to 2013-09-16 UTC.
Each snapshot's AOR produced 14 subarray FITS files containing 896 individual
images.  Continuous staring mode observations were made as 5 roughly 12
hour long, consecutive AORs from 2012-04-24 to 2012-04-26 UTC.

For HD~158460, a 0.1s frame time was chosen to observe this 1187~mJy
calibration star at $\sim32\%$ of full well, where the IRAC detectors
have the least non-linearities.  In total, 18 snapshot observations
were taken randomly from 2012-01-13 though 2012-01-20 UTC. Each
snapshot AOR produced 210 subarray FITS files containing 13,440 individual
photometry measurements.

We do not use the standard 30-min pre-AOR as that would remove the
efficiency gain provided by using snapshots over continuous stares.
The goal of that pre-AOR is to allow the telescope motion to settle
upon first arriving on target so that the target does not drift off of
the sweet spot.  Since our observations are only 30 minutes long, as
opposed to the many hour continuous stares, the drift in position that
occurs within the observations keeps the majority of the snapshots
within the calibrated region of the pixel, so the pre-AOR is not
necessary for this particular application.

Four of the 51 AORs did not end up with at least 20\% of their
datapoints on the sweet spot.  This includes one continuous staring
mode AOR (containing the second observation of the secondary eclipse)
and three of the snapshot AORs.  These non-sweet-spot AORs are ignored
in the following analysis.

\section{Data Reduction}
\label{data}

Data reduction is exactly the same for all snapshot and continuous
staring mode datasets.  We start our data reduction from the Basic
Calibrated Data (BCD) files provided by the SSC data pipeline,
software version S19.1.0.  The pipeline applies a dark subtraction,
linearization, flat-field correction, and conversion to flux units.
The first subframe of each 64-frame BCD is removed from this analysis
due to its higher bias level.  The higher bias level is caused by the
larger elapsed time between BCDs as opposed to consecutive
subframes within a single BCD.  This effect is referred to as the
``first frame effect'' in the IRAC documentation
(\url{http://irsa.ipac.caltech.edu/data/SPITZER/docs/irac/iracinstrumenthandbook/}).

\subsection{Superdark}
\label{superdark}
Dark current in all IRAC frames is measured with a shutterless system
designed by the IRAC instrument team \citep[described in detail in ][]{2009ApJS..185...85K}.  In
summary, weekly observations at each frame time for both sub- and
full-array are made of a low background field at the
north ecliptic pole.  These images are then vetted for obvious
persistent images (latents), and median combined to make a weekly
dark.  For each data frame, the nearest-in-time dark is used by the
pipeline to correct that frame.  This means that for our datasets
which span multiple weeks/years, different dark frames are subtracted
from different AORs. For our science goals, we need to put aperture
photometry from the whole set of snapshot observations on a single
flux level.  To determine how best to do this, we examine the effect
of using weekly darks in comparison to a mission-long superdark.  The
reason for concern is that a latent image in the dark frame near the
center of the frame (either in the aperture or the background annulus)
will cause changes in the aperture photometry of the target from one
AOR to the next.

Because the latent population changes on day to week long timescales,
the advantage of weekly darks is that they potentially have the same
latent in them that the science data has, allowing us to remove
latents from the science frames.  However it is also possible that the
weekly darks have latents in them which have faded by the time the
science data are observed, thereby adding a source of noise to the
science data.  It is possible to generate a superdark from
archival data by median combining all dark images taken during the
entire duration of the warm mission to date.  Each frame time gets its
own superdark.  The advantage of superdarks is
that they have no latent structure in them.  The disadvantage of
superdarks is that they cannot remove latents which are present in the
science data.

When making both the weekly and superdarks, the median combine will
reject all astronomical sources.  Aside from latent images, we expect
that the only other difference between the superdark and weekly dark
to be a change in the background level because the zodiacal light
component varies as a function of time \citep{2012ApJ...754...53K}.
However, since we are doing aperture photometry, we do not care that
the mean level of the dark is incorrect by the amount that the
zodiacal light fluctuates over a one-year baseline.  Throughout the
warm mission we do not see evidence of a change in the background
pattern in the darks.  The superdark is applied to each data frame by
first backing out the pipeline calibrations already applied to the BCD
including the weekly dark, and then calibrating those images with our
superdark.  All of the calibration files required to do this are
provided by the SSC in the Spitzer Heritage Archive.

Without the luxury of darks taken directly adjacent to the snapshots,
we are forced to choose between using the weekly darks and the
superdarks.  We use the standard deviation within datasets as the
metric for the decision of weekly darks versus superdarks.  We find
the lowest scatter photometry using weekly darks for the gain map
calibration dataset and superdarks for the WASP-14b dataset.  This is
probably due to the persistent images we see in the snapshot dataset
(see \S \ref{outliers}), whereas the gain map dataset is so much
larger that the individual frames have not been studied in such
detail.

\subsection{Centroiding and Aperture Photometry}
\label{cent_phot}
The centers of the star on each subframe are determined using the SSC
provided box\_centroider.pro
(\url{http://irachpp.spitzer.caltech.edu/page/contrib}).  This code
uses an iterative process with the first moment of light to find the
star centers. We choose this technique because it is simple, robust,
repeatable, and easy to code for comparison with other groups using
different methods.  We have not tested other centroiding methods on
this dataset.  For WASP-14b, there is a faint star $\sim11.4$
arcseconds from the center of WASP-14 (not cataloged in Simbad so we
don't have literature fluxes for it).  A $6 \times 6$ arcsecond region
around the right ascension (RA) and declination (dec) of that star is
masked in all images to block the light coming from that star before
centroiding and performing photometry.  The masked region is well
outside of photometric aperture, but does cover part of the background
annulus.

Aperture size is chosen for this dataset by finding that aperture
radius which maximizes the SNR of the final reduced dataset.  We test
eight apertures in 0.25 pixel bins from 1.5 to 3.25 pixels.  For
WASP-14b, a 2.25 pixel aperture minimizes the contribution from
background noise, while including enough of the star's flux to
maximize the SNR.  Since the calibration star HD~158460 is a test
dataset, we hold the aperture size constant at 2.25 pixels for that
dataset as well.  We use this same aperture size on the gain map
dataset for consistency.

A background annulus of 3 - 7 pixels (3.6 - 8.4 \arcsec) is used.  We
choose this annulus size for WASP-14b because there is a neighboring
object in the images at about 10 pixels from the center of WASP-14,
and since we have no information on the variability of that source, we
would like to exclude it from our background region as much as
possible.  Also, since we need to use a single, uniform method for finding the
background on our target as well as on the very large gain map
dataset, we have chosen to stick with the 3-7 pixel annulus for the
entire paper.

Along with centers and aperture fluxes, we calculate the number of
noise pixels for each subframe as well as the individual components of
noise pixel in the X and Y direction, calculated using the SSC
provided box\_centroider.pro. The noise pixel parameter gives an
indication of apparent size of the target star and is defined in the
IRAC Instrument Handbook
(\url{http://irsa.ipac.caltech.edu/data/SPITZER/docs/irac/iracinstrumenthandbook/5/})
and in \citet{2005MNRAS.361..861M}\citet{2013ApJ...766...95L}.  Specifically it is the equivalent
number of pixels whose noise contributes to the flux of a point
source.  There is no evidence that the PSF itself is changing with time.
However, at frequencies higher than the observation rate, oscillations
of the spacecraft will have the effect of smearing out the image,
thereby increasing noise pixel, without changing the centroids
significantly.

To bin, we remove both position and
flux outliers.  We first remove points from our dataset that are three
standard deviations away from the global mean $Y$ position or 2.5
standard deviations away from the global mean $X$ position.  The
larger tolerance in the Y direction is due to more native spacecraft
motion in the Y direction.  We have checked individually many of these
position outliers, and every one we have followed up is caused by a
cosmic ray near to the star which corrupts the measured positions. For
both the snapshot and continuous staring datasets we reject on average
$\sim1\%$ of the data as position outliers.  Flux outliers are removed
using a running three sigma mean.

\begin{figure*}
\hspace*{-1.5cm}
\includegraphics[scale=0.5]{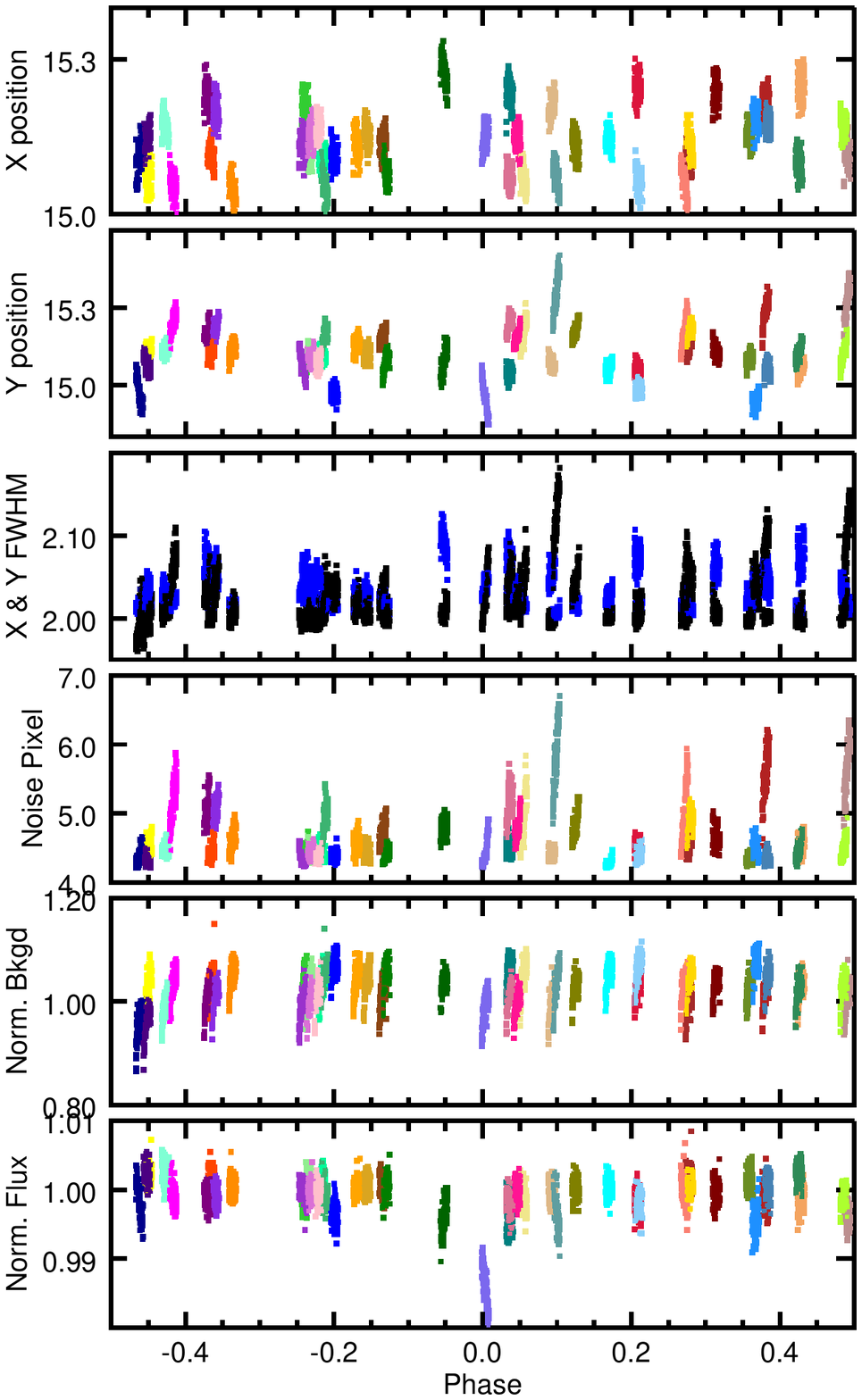} 
\hspace*{-2.5cm}
\includegraphics[scale=0.5]{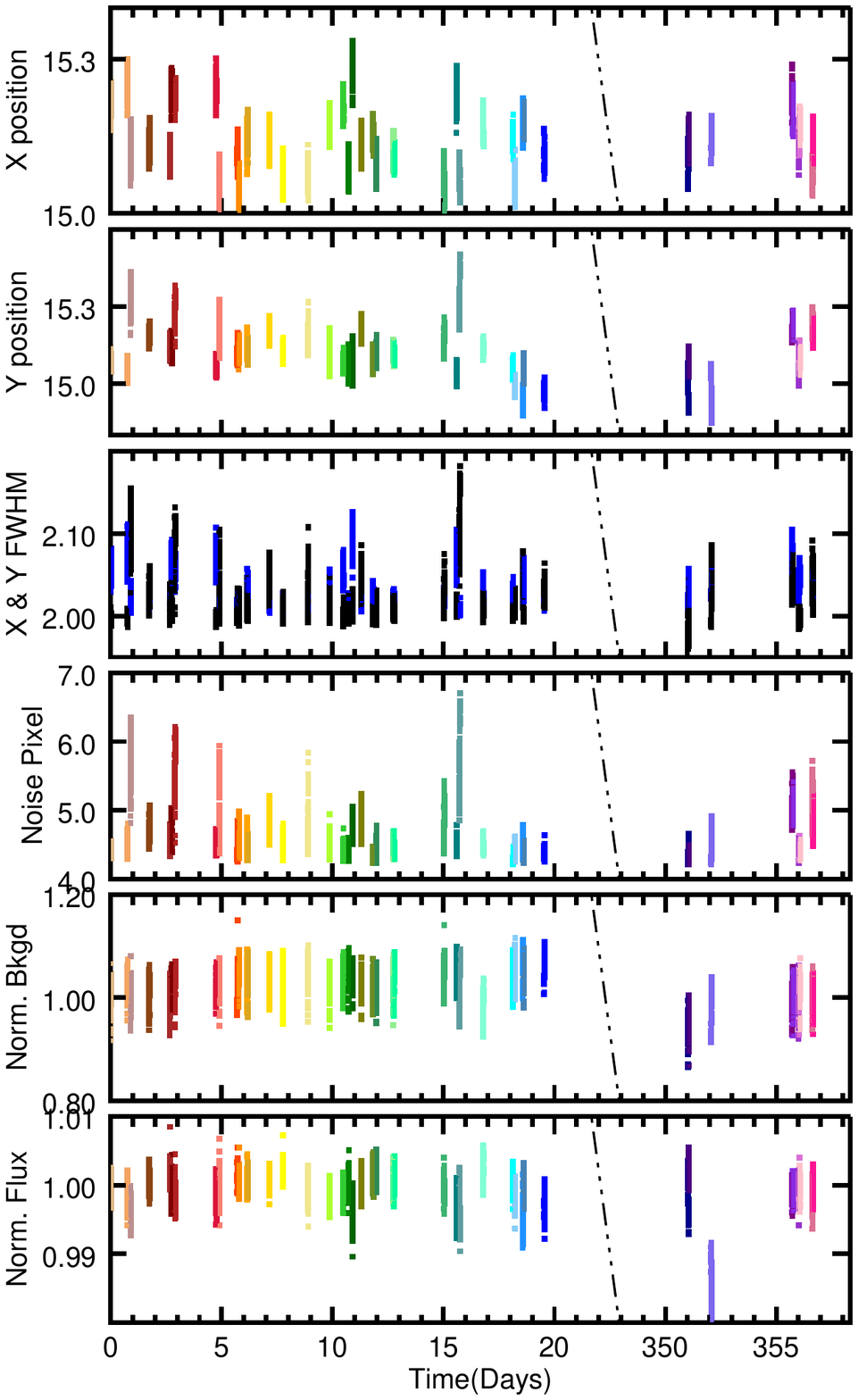} 
\caption[multiplot]{ WASP-14b raw data as a function of phase on the left and
  time in days on the right.  Each AOR is
  color coded with the exception of X and Y FWHM where XFWHM is shown
  in blue and YFWHM is shown in black.  All the data are in reasonable
  ranges of the important variables shown.  The time plot has a
  discontinuity of about a year between the first 35 and the last 11
  AORs shown by the diagonal dot-dashed lines.  See \S \ref{cent_phot}
for more details.}
\label{fig:multiplot}
\epsscale{1}
\end{figure*}

\begin{figure*}
\hspace*{-1.5cm}
\includegraphics[scale=0.5]{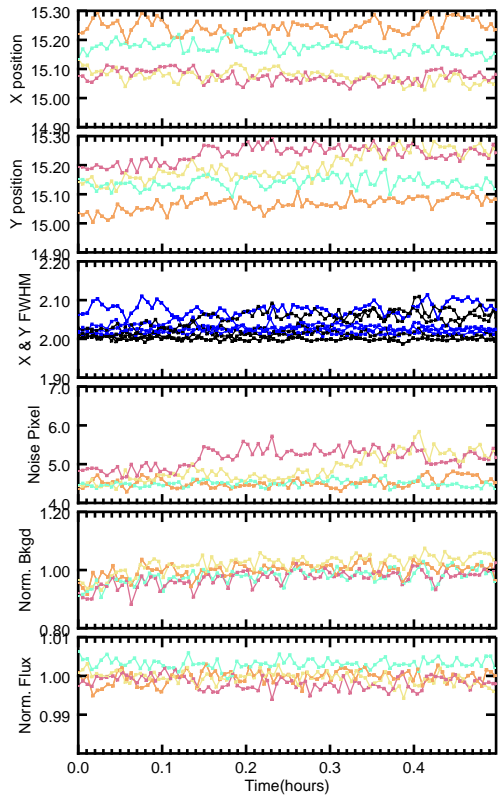} 
\caption[multiplot_few]{ A zoom in on WASP-14b raw data as a function
  of time for four random AORs spread throughout the set of observations.  Each AOR is
  color coded with the exception of X and Y FWHM where XFWHM is shown
  in blue and YFWHM is shown in black. .  See \S \ref{cent_phot}
for more details.}
\label{fig:multiplot_few}
\epsscale{1}
\end{figure*}

\begin{figure}
\hspace*{-1cm}
\vspace*{-2cm}
\includegraphics[scale=0.5]{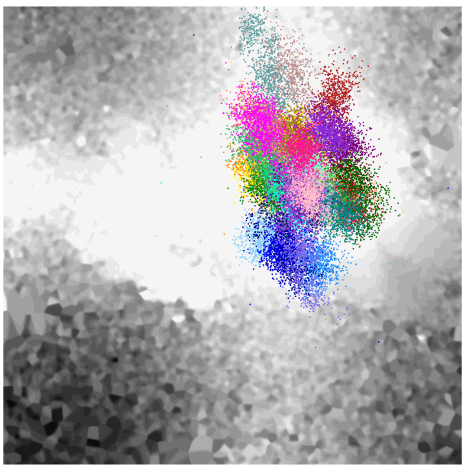}  
\caption[Position]{ Greyscale image of the gain variations within a single pixel
  at the center of the subarray.  The centroid positions of WASP-14b snapshot
  AORs are shown in the same color as the other WASP-14b figures. See \S \ref{cent_phot}
for more details.}
\label{fig:position}
\epsscale{1}
\end{figure}

Figure \ref{fig:multiplot} shows the raw values of $X$ centroid, $Y$ 
centroid, X \& Y FWHM (second moment of intensity distribution), noise
pixels, normalized background value, and normalized raw flux, as a
function of both orbital phase and time.  The time plot includes a
discontinuity of about a year between sets of observations.  We also
show a zoom in for a few AORs on Figure \ref{fig:multiplot_few}.  Figure
\ref{fig:position} shows all the snapshot AOR centroid positions
over-plotted on an image of the intrapixel gain map to give a sense of
how accurate and precise the pointing is for this program.  The color
coding of the AORs remains consistent throughout all figures for this
paper.

\subsection{Kernel Regression Gain Map Correction}
\label{hybrid}
Data for both the snaps and stares were reduced using a
Nadaraya-Watson type kernel regression technique
\citep{Nadaraya:2006de,Watson:1964kb}. Kernel regression has been used
extensively in the literature as a way of correcting for the IRAC
intrapixel gain effect based on neighboring photometry points in the
dataset itself (often referred to as pixel mapping)
\citep{2010PASP..122.1341B,2012ApJ...754...22K,2013ApJ...766...95L,2014ApJ...790...53Z,2015ApJ...811..122W}. Different
from the literature methods, however, we use the photometry of a
separate calibration star BD+67 1044, which is not known to vary.  As
mentioned earlier, the {\sl Spitzer} Science Center has accumulated
approximately 400,000 measurements of this gain map calibration star,
positioned all within 0.3 arcseconds of the ``sweet spot'' (peak of
response) of the ch2 subarray central pixel.

Kernel regression correction of science data begins by finding the $N$
nearest neighbors to a given target data point in the gain map
calibration dataset, based on the Euclidean distance in $X$ and $Y$
centroid and noise pixels. The $N$ gain map points are weighted by a
kernel which is a Gaussian function of the distance to the science
data point.  The width of the Gaussian kernels are computed on the fly
as the coordinate standard deviations (in x,y,NP, etc) of the 50
neighbors in the pmap dataset.  The weighted kernels are then summed and
normalized by the calibration star flux.  The result is a prediction
of the relative strength of the intrapixel gain map at the location of
the science data point. To correct for the intrapixel gain we divided
the science flux by this prediction.

The technique includes tunable parameters for the number of nearest
neighbors $N$, the maximum distance from the science data point, and the
minimum ``occupation'' number, a kernel-weighted measure of how many
of the nearest neighbor gain map data points contributed to the result
(see \cite{2012SPIE.8442E..1YI} for details on the occupation).  For
the current program, we set the number of nearest neighbors to 50 \citep{2013ApJ...766...95L}, the
minimum distance to 0.0025 pixels, and
the occupation number to 20.  This tight requirement on the proximity
and quantity of gain map data helps to minimize the effects of
persistent images that may exist in the pmap dataset. As mentioned
above, we did not include three AORs in our final analysis that have
less than twenty percent of their photometry points within the sweet
spot. One important difference between this method and literature
kernel regression techniques for exoplanet reduction is that we did
not build a correction from the science measurements themselves, and
therefore minimize the risk of removing astrophysical signal.

An earlier version of this technique contained an intermediate step of
computing the gain map on a regular grid in $X$ and $Y$ centroid, and
then interpolating to the science data positions
\citep{2012SPIE.8442E..1YI}.  We find the direct nearest neighbors
approach to be more effective (albeit more CPU-intensive), because
working from an intermediately gridded map introduces additional
uncertainties inherent to grid-ding and interpolation.  The current
method will be described more fully in a future paper (Ingalls et al,
in preparation).

\begin{figure}
\hspace*{-1cm}
\vspace*{-4cm}
\includegraphics[scale=0.7]{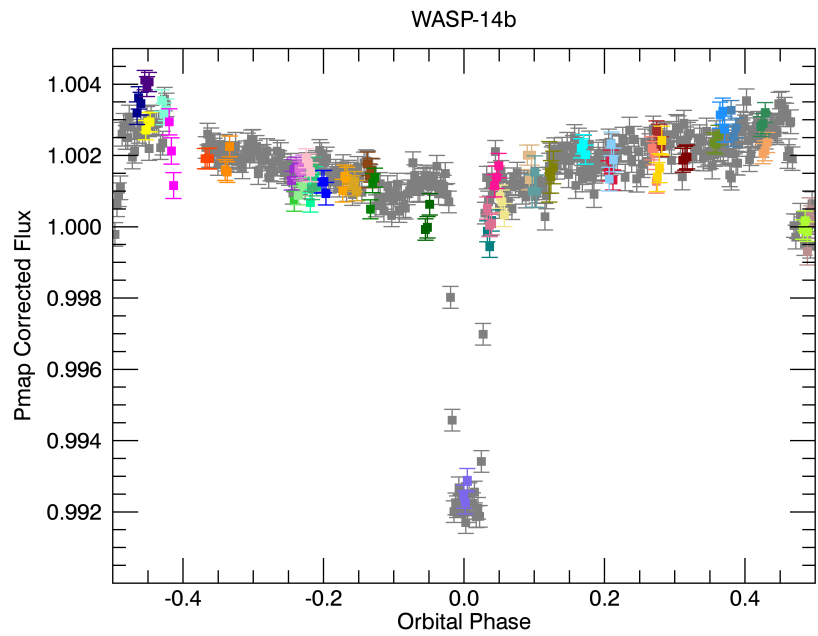}  
\caption[WASP-14b]{Final Photometry of WASP-14b.  Gray points show our
  reduction of the continuous staring mode data. Colored points show the snapshot
  data.  All have been binned on the same timescale and normalized to
  the level in secondary eclipse. Zoom-ins on the transit and eclipse
  are shown in Figure \ref{fig:secondary}.  See \S \ref{hybrid} for
  more details.}
\label{fig:WASP-14b}
\epsscale{1}
\end{figure}

\begin{figure*}
\hspace*{-0.5cm}
\includegraphics[scale=0.4]{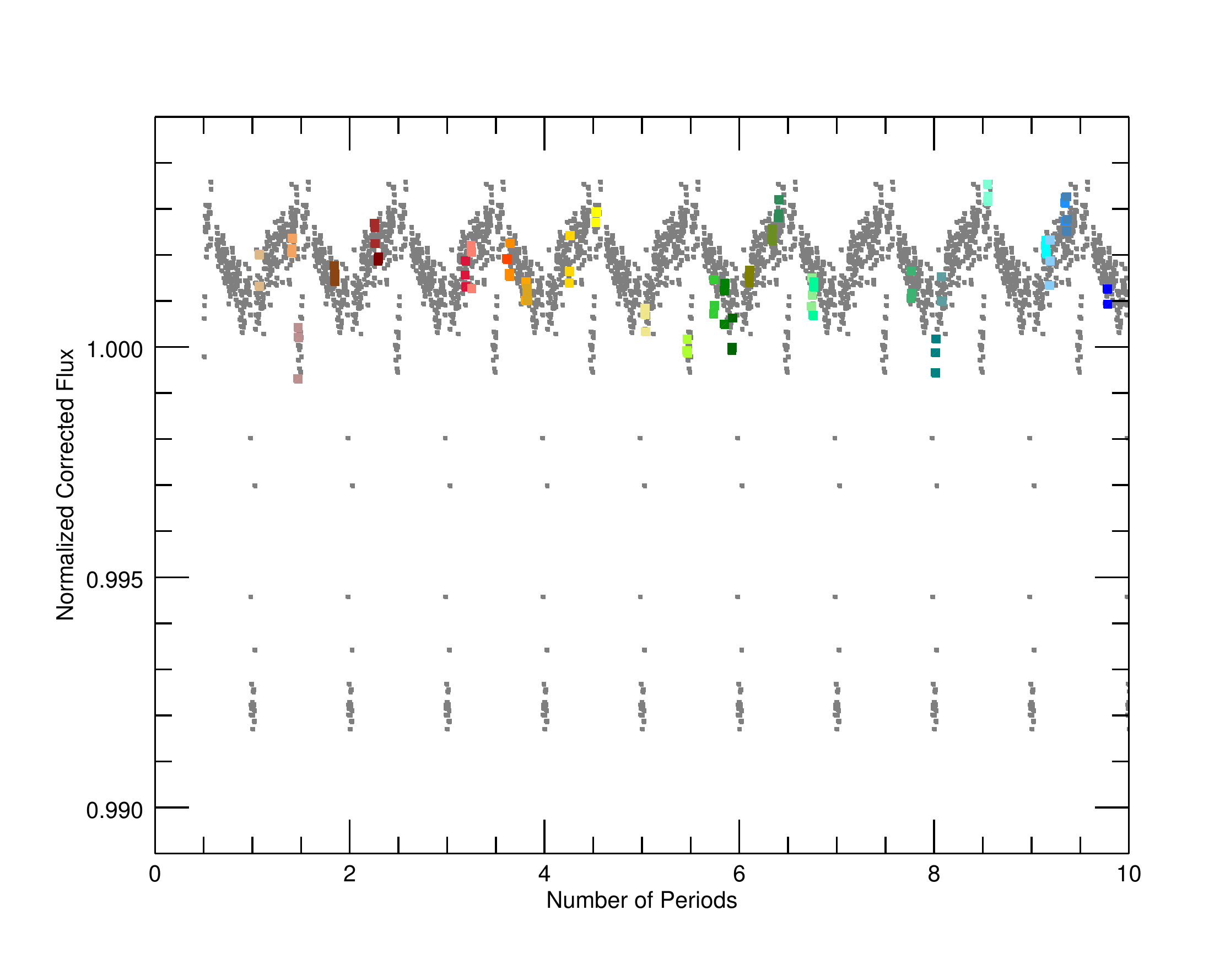}
\hspace*{-0.5cm}
\includegraphics[scale=0.4]{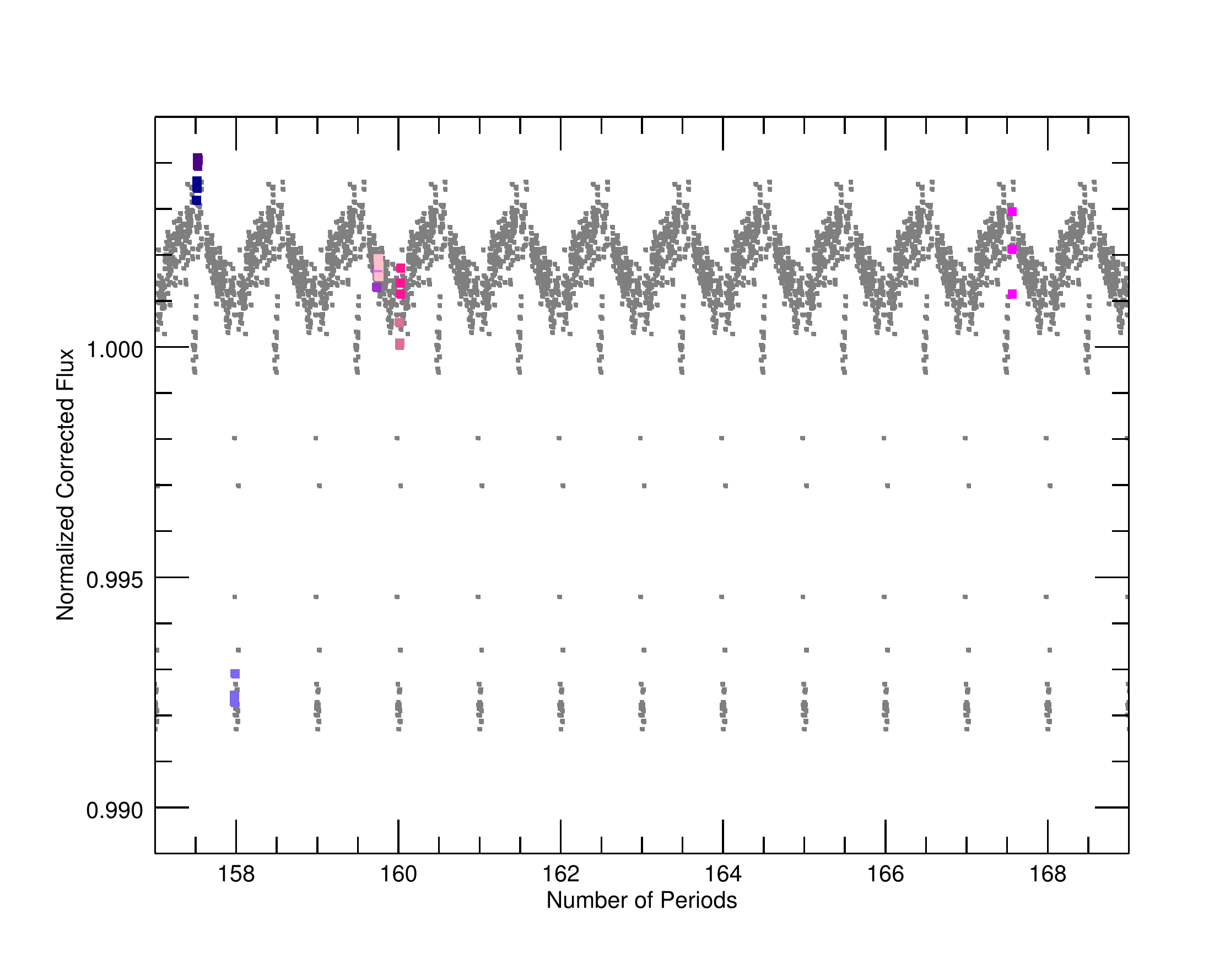}  
\caption[flux_period]{ Flux vs. Time over the 1 year of snapshot
  observations.  The x-axis is in number of periods since the first
  snapshot observation.  Gray points are phased continuous staring mode
  observations repeated over many periods.  Colored points are the
  snapshot data.  Binning is the same for all data. See \S \ref{hybrid} for
  more details.}
\label{fig:flux_period}
\epsscale{1}
\end{figure*}

Figure \ref{fig:WASP-14b} shows the final reduction of WASP-14b
snapshots and continuous staring mode AORs phased to the literature period 
of 2.24376507 days \citep{2015ApJ...811..122W} . Figure
\ref{fig:flux_period} shows the same data as a function of time
instead of phase. Since there is about one year between each set of
observations, we show this in two frames with the first set of 35 AORs
in one frame, followed by the second set of 11 AORs in the other.

\subsection{Nonlinearity}
\label{nonlin}
The IRAC detectors have a known nonlinearity such that a linear
increase in the number of incoming photons does not lead to a linear
increase in the flux measured on the detectors.  The pipeline corrects
for this nonlinearity to an accuracy of about 1\%.  High precision
photometry studies which want to use the gain map taken at one flux
level to correct science data taken potentially at another flux level
are sensitive to a 'residual nonlinearity'€ ($<1\%$).  We have examined this
residual nonlinearity in detail \citep{IRACHPP} and conclude that it
does not effect flux as a function of position on the pixel for
observations with peak pixel counts between 1,000 and 15,000 DN in
ch2, which is the case for all observations here.  WASP-14b has
between nine and ten thousand DN in the peak pixel depending on the
position.

\subsection{Electronic Ramp}
\label{ramp}
\begin{figure}
\hspace*{-1.5cm}
\vspace*{-3.5cm}
\includegraphics[scale=0.5]{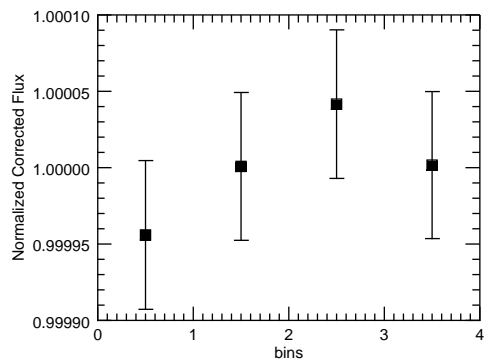}  
\caption[ramp_check]{Corrected flux as a function of time.  Fluxes for
  all 46 snapshot AORs have been binned together.  Time is binned into
  4 sections to look for trends as a function of time.  Y-scale range
  is 0.02\%.  There is no electronic ramp from the beginning to the end
  of the snapshot observations to within 50-100ppm, which is well
  below our phase curve amplitude signal. See \S \ref{ramp} for more details.}
\label{fig:ramp_check}
\epsscale{1}
\end{figure}

Some authors report an electronic ramp in flux measurements as a
function of time in their warm IRAC datasets
\citep{2011ApJ...727..125C,2011ApJ...726...95D,2012ApJ...746..111T,2013ApJ...770..102T}.
Ramps seen in the 5.8 and $8.0\micron$ cryogenic data are probably not
relevant because those are Si:As detectors whereas the ch1 \& 2
detectors are InSb (see
\url{http://irsa.ipac.caltech.edu/data/SPITZER/docs/files/spitzer/preflash.txt}
for a discussion of the $8\micron$ ramp).  We see no evidence for a
consistent ramp in our 30 minute AORs.  Figure \ref{fig:ramp_check}
shows normalized corrected flux as a function of time for all 46
snapshot AORs binned together into 4 bins as a function of time.
There is no systematic trend from the beginning to the end of the
observations within 0.005\%, which is well below the noise level of
our phase curve measured amplitude.

\section{Results \& Discussion}
\label{results}

We first examine persistent images as a remaining noise source after
the above reduction.  We then fit a model to our reduction of
both the snapshot and continuous staring mode WASP-14b data.  Next, we consider
three independent metrics for judging the success of the snapshot
technique.  Finally we look at the astrophysical implications of the
WASP-14b dataset.
\subsection{Residual Persistent images}
\subsubsection{Outliers in Snapshot Data}
\label{outliers}
The existence of persistent images under the snapshots is likely the
largest uncorrected source of scatter in our snapshot dataset.  We
examine the effect of persistent images on the snapshot observations
(see \S \ref{latent}).  Our only metric for determining which of the
AORs are effected by stronger latent signals is to reduce the data in
two ways and then compare; first using a superdark (theoretically no
latents) and secondly using weekly darks.  Out of 46 AORs we find only
three AORs whose flux densities change by more than 0.05\%: the two
AORs directly after the first secondary eclipse (purple and dark blue)
and the AOR during transit(light purple).  These three AORs were taken
consecutively in time, implying that latents are the likely source of
scatter in those AORs.  Conversely we can say that the remaining 43
snapshots are not affected by an amount greater than 0.05\% by latent
images.  This is confirmed both by the small amount of scatter in the
residuals (see Section \S\ref{rms}) and in the difference between
reductions with and without the superdark.

\begin{figure*}
\hspace*{-1cm}
\includegraphics[scale=0.5]{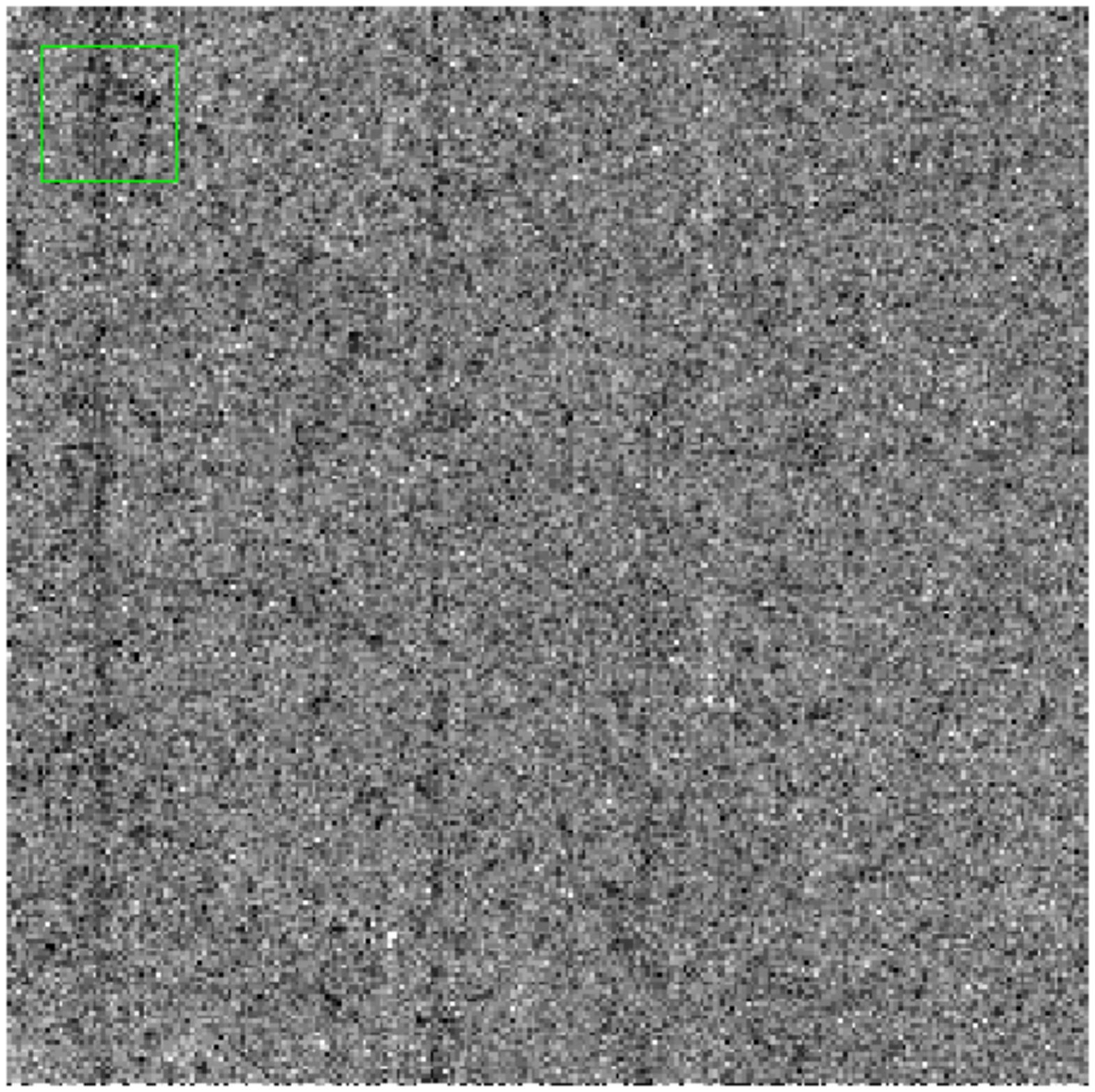}  
\hspace*{-1cm}
\includegraphics[scale=0.5]{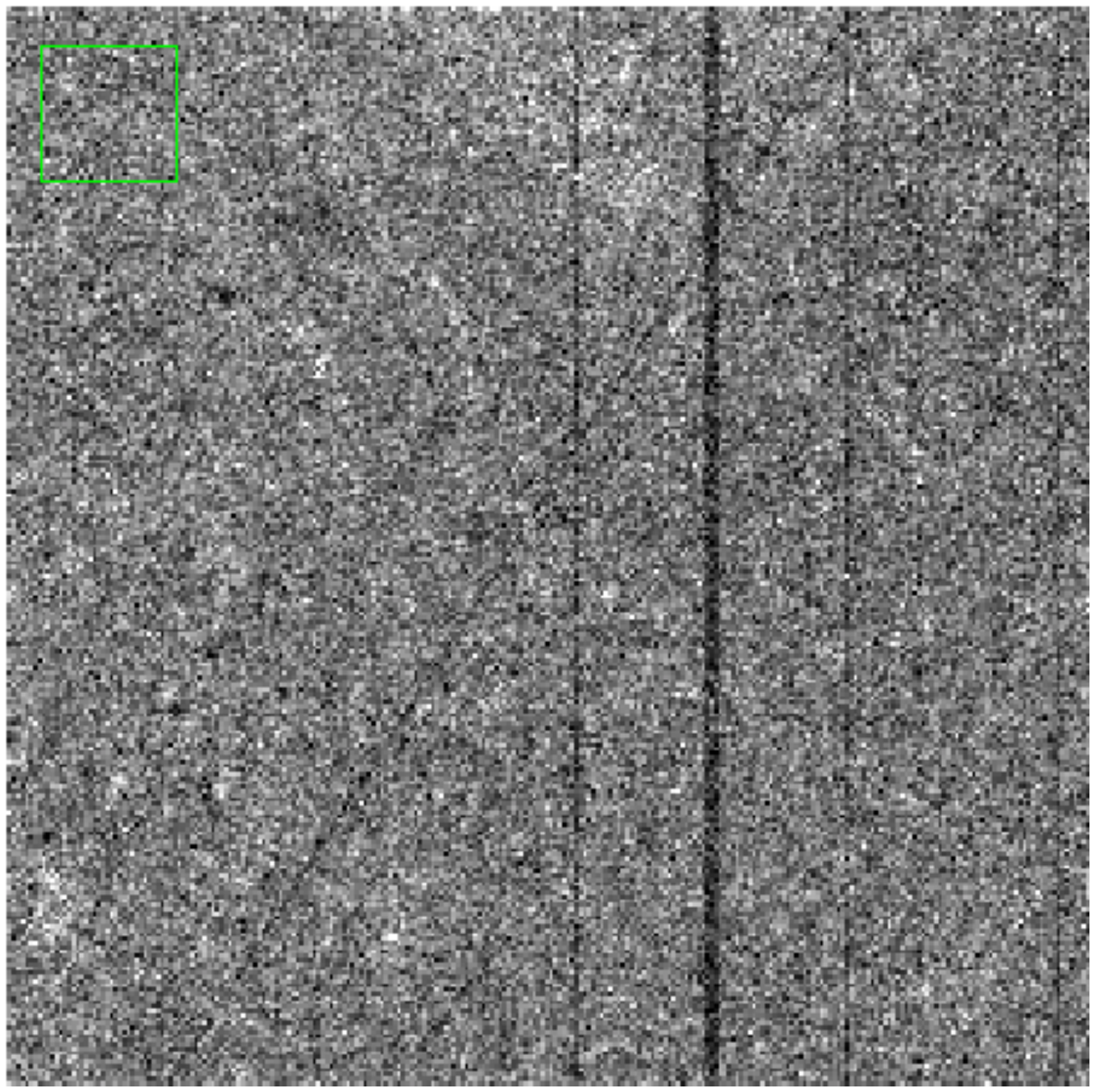}  
\caption[latent]{Left: Median Stack of 80 images taken after a set of
  snapshot observations with full array frame time of 6s.  Right:
  Median stack of 10 images taken after a different snapshot with full
  array frame time of 0.4s.  The green squares show the location of
  the subarray field of view.  There are clearly negative (more darkly
  colored) persistent images in the subarray field of view in one of these median stacks
  and not obviously in the other.  The darkly colored vertical stripes
  are persistent images from a column-wise effect known as column
  pull-down.  The diagonal latent in the right side of the figure is
  a slew latent where the telescope must have slewed across a bright
  star at some time prior to these observations.  See \S \ref{outliers} for more
  details.}
\label{fig:latent}
\epsscale{1}
\end{figure*}

To further test if AORs with the largest delta flux density from
superdark to weekly dark are indeed affected by latents, we first
looked for a source of persistent images observed directly prior to
the candidate high latent snapshot AORs.  Approximately 2.5 hours
prior to the start of the snapshot observations, IRAC mapped the Chameleon
region, using a dithering and mapping strategy and not staring.  This
region includes some bright stars with $K$ = 3-5 that could cause
low level persistent images. Furthermore, we examined an AOR taken
directly after the three snapshot AORs to look for residual persistent
images. This AOR is fortunately 80 BCDs that dither around a
relatively dark portion of the sky. Median combining these images
together removes all sources, and leaves us with an image of the
background pattern which may have been present underneath our snapshot
observation ( see left side of Figure \ref{fig:latent}). Indeed there
are persistent images present in the median combine both from column
pulldown, and potentially from other bright targets all over the full
array, including in the subarray region (small box in upper corner).
The subarray persistent image is potentially the source of the
different flux in the three snapshot observations examined here.

For comparison, we also look at median stacks of AORs observed
directly after two other random snapshots (first, and 27th snapshots), and these
show a smooth background in the subarray region, albeit with lower SNR
in the background due to less total exposure time in the median
stack (see right side of Figure \ref{fig:latent}).  

Not only are there long term persistent images, as seen in the
superdark analysis, but there are also short-term persistent
images which decay much faster (minutes) which can affect the
photometry at the 0.1\% level.  Additionally it is possible that the
exoplanet targets themselves are also generating persistent images,
but we cannot disentangle this effect in the source photometry.
Our suggestion above of dithered observations before and after each
snapshot will allow us to study these types of latents and improve
future observations.  For the three AORs that are affected by latents,
we have added 0.1\% to their error bars to more accurately represent
the uncertainty in flux due to persistent images.

\subsubsection{Level Offset}
One side-effect of the persistent images underneath all of our
observations is that the continuous staring mode data and the snapshot
data have different normalizations.  We can not know exactly what the
latent behavior under these observations is, but we know that the
latent behavior is different from season to season and year to
year. Because this is an additive effect, we correct it by adding an offset of
0.2\% uniformly to all snapshot data to equalize the snapshot mean
flux with the continuous staring mode data (judging by eye since both sets of
observations have scatter and phase curve shape). Since we are not
doing absolute photometry, the additional flux offset that we manually
add will make minimal difference in the conclusions of the paper. 

\subsection{Phase Curve Fitting}
\label{fitting}

To test the usefulness of snapshot data sets for deriving physical
parameters, we fit the \citet{2013ApJ...766...95L} phase curve model
to the observed phase curve of WASP-14b.  This model is based on a fit
of sines and cosines for circular orbits presented in \citet{cow08}
but has been adjusted to accommodate systems with eccentricity.  Given
a full, continuous, phase curve, there would be up to 16 fitted
parameters: orbital period, inclination, $a/R_*$, two components of
eccentricity, transit mid-point, $R_p/R_*$, depth of both secondary
eclipses, four phase curve parameters, and two ramp parameters.
However, since the snapshot style observations were not designed to
measure information about transits and eclipses, and in fact those are
not guaranteed to be observed in any given set of snapshots, we fix 12
of those parameters and allow only the four phase curve shape
parameters to vary.  The 12 fixed parameters are fixed to the values
presented in \citet{2015ApJ...811..122W}, listed in Table \ref{tab:fitpars}.  The phase curve shape is
defined by the following equation:
$$F(\theta) = F_0 + c1cos(\theta) + c2sin(\theta) + c3cos(2\theta) +c4sin(2\theta)$$

\noindent where c1-c4 are the free parameters, F is flux and $\theta$ is the
phase which, in the case of an eccentric orbit,  is a function of the
true anomaly \citep{2013ApJ...766...95L}.

\begin{figure}
\hspace*{-0.5cm}
\includegraphics[scale=0.6]{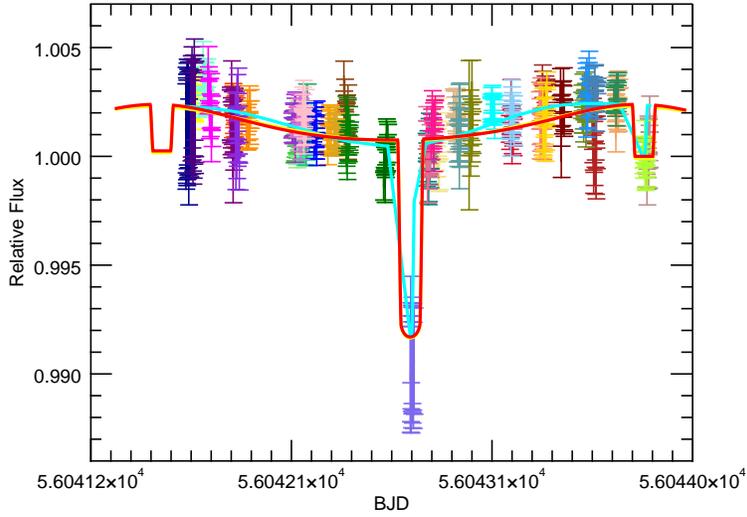}  
\caption[fit_latenterr]{Fits to the snapshot and continuous staring
  mode datasets.  Data are binned roughly by 64 photometry points. The
  red line is the fit from this paper to the pmap reduction of the
  continuous staring mode data, the cyan line is the fit to the
  snapshot data, and the yellow line is the
  \citet{2015ApJ...811..122W} fit to their reduction of the continuous
  staring mode data(practically indistinguishable from the red line).
  The cyan line fit to the snapshot data is slightly irregular at
  transit and eclipse because there isn't enough data in ingress and
  egress to provide a smooth fit.  An extra 0.01\% has been added to
  the error bars of all data in the AORs which we know are effected by
  latent images(most notably those in transit). See \S \ref{fitting}
  for more details. }
\label{fig:fit_latenterr}
\end{figure}

Figure \ref{fig:fit_latenterr} shows the snapshots and the best fits for both the
continuous staring mode and snapshot data which are both reduced in
exactly the same manner using the gain map.  Additionally, we over-plot
the fit from \citet{2015ApJ...811..122W} to their reduction of the
continuous staring mode dataset. Data are shown binned at the same
level at which the fitting is performed.  We found the best chi-squared
values for binning at roughly the level of a set of 64 photometry
points (equivalent to one fits file).

After fitting, values of phase curve amplitude and phase shift are
derived.  Uncertainties on these values are calculated by running a
set of 1000 fits on a dataset where the fitted parameters are randomly
varied within their $1\sigma$ error bars.  The distribution of
resulting amplitudes and phase shifts is then fit with a gaussian and
the uncertainties are obtained from that gaussian fit.  Error bars in
the pmap fits are likely underestimates.

Maximum flux occurs prior to phase = 0.5 and likewise minimum flux
occurs prior to phase = 0.  This is fully consistent with
\citet{2015ApJ...811..122W} for their continuous staring mode
analysis.   It implies that the hot spot is shifted eastward
from substellar point, which is found for other hot Jupiters as well
\citep[eg.][]{knu07,2012ApJ...754...22K,2014ApJ...790...53Z,2014Sci...346..838S}.
General circulation models predict that the eastward hot spot implies
an eastward super-rotating equatorial jet stream caused by the day to
night temperature differential \citep[][and references
therein.]{2015ApJ...801...95S}.

\subsubsection{Comparison with Continuous Staring Mode Data}

We compare our best fit model for the snapshot data with two models
from the continuous staring mode data.  Those two models are 1) our
own reduction and fit of the continuous staring mode data and 2) the
independent \citet{2015ApJ...811..122W} model fit to their reduction.
Figure \ref{fig:fit_latenterr} shows all models,
\citet{2015ApJ...811..122W} in yellow, this paper's continuous staring
mode model in red, and the snapshot model in light blue. Our model fit
to our own reduction of the continuous data and the
\citet{2015ApJ...811..122W} fit to their reduction are extremely
similar, validating this fitting technique.  All three
models find similar values of phase shift within the uncertainties.

The derived value of amplitude is statistically different
($\sim 5\sigma$) between the snapshot fit and the continuous data
fits.  Our fitting routine does find a slightly higher amplitude than
that published in \citet{2015ApJ...811..122W}, but it is consistent
with their value.  We test if this is a function of the sampling into
snapshots by subsampling the continuous data into snapshot style
observations, and fitting those observations.  We do this 100 times,
and find a distribution of amplitudes that is consistent with our
fitting of the continuous data.  Therefore, we conclude that the
sampling of the phase curve into snapshots is not the cause of the
difference in amplitudes.

Rather than assume there is an astrophysical source of this
discrepancy, we assume instead that the uncertainties on our
photometry points are underestimated because they do not include a
contribution from the persistent images.  Persistent images varying
between snapshot observations (see \S\ref{outliers}) will cause the snapshot
observations to vary from one epoch to another, which can cause
changes to the measured amplitude of the phase curve.  Increasing the
uncertainties on the snapshot photometry by 0.0001 brings the measure
of phase amplitude in the snapshot to within the uncertainties of
those measured for the continuous dataset.  This is consistent with the level
as estimated above in our search for persistent images in some of the
observations after our snapshots.


\subsection{Do Snapcurves Work?}

We present two additional lines of evidence that the snapshot strategy can
successfully obtain phase curves.  First, we look for residual
time correlated noise in our snapshot observations.  Second, we look at snapshot
observations of a calibration star which should recover a flat light
curve. 

\subsubsection{Time Correlated Noise}
\label{rms}
\begin{figure}
\includegraphics[scale=0.7]{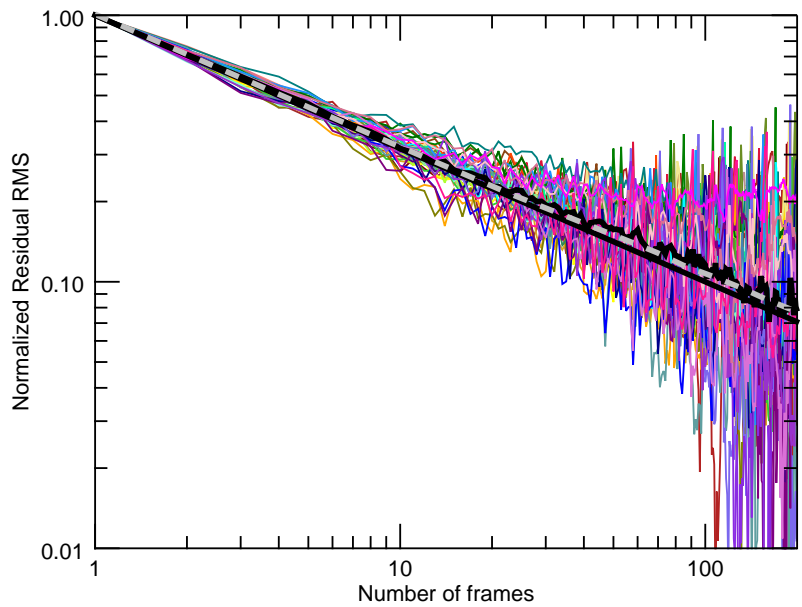}  
\caption[rms]{ Normalized Residual RMS versus binning size in number
  of frames on the bottom and minutes on the top.  Each colored line
  represents a single AOR.  Because our observations are only 30 min
  long, the RMS measurement has a lot of scatter in it at binning
  scales greater than a few minutes.  The straight solid black line is poisson
  noise.  The gray dashed line is the same measurement for the continuous staring mode
  data taken from \citet{2015ApJ...811..122W}.  The squiggly black
  line is the median over all snapshots.  See \S \ref{rms} for
  more details.}
\label{fig:rms}
\epsscale{1}
\end{figure}

To test how well our reduction technique removes sources of correlated
noise, we compare the rate at which binning reduces the noise to the
expected rate of square root of the number of data points for strictly
poisson noise.  Figure \ref{fig:rms} shows this comparison for each of
the snapshot AORs.  To generate the Y-axis, we subtract the (fully
independent) \citet{2015ApJ...811..122W} model from each of our
snapshot data, and measure the RMS of the residuals.  We then bin the
data on increasingly large scales and plot the results as a function
of number of frames per bin.  Because we only have 30 minute long
AORs, these plots show significant scatter as the total number of bins
per AOR gets small. The solid black line is the expected binning
relation for poisson noise.  The dashed grey line comes from the
\citet{2015ApJ...811..122W} reduction of the WASP-14b continuous
staring mode dataset.  This continuous dataset has a much longer
duration and is therefore able to probe larger binning scales.  The
median of all snapshot AORs is shown with the black squiggly line, and
similar to the \citet{2015ApJ...811..122W} reduction, is within about
10\% of Poisson noise.  For reference, our own reduction of the continuous staring
mode data also follows the \citet{2015ApJ...811..122W} slope.

The bright pink points at phase -0.4 in the light curve plots which
appear more extended in corrected flux as a function of phase and time
also diverge strongly from poisson slope in this plot (flatten out to
a normalized residual RMS of 0.2 by 20 binned frames).  The cause for
the increased scatter is unknown.  The observations taken after this
particular AOR are too short, and include bright stars, to determine
if there could be an underlying rapidly-varying latent.  There is
nothing obviously unusual about the position, noise pixel or
background level in that AOR as seen in Figures \ref{fig:multiplot}
and \ref{fig:position}, so we do not suspect the gain map correction.

Figure \ref{fig:rms} shows that the nearest neighbor gain map method
presented in this paper is 1) consistent with other reduction methods
in that we have adequately removed the intrapixel gain effect,
2) multi-epoch data has not added new sources of correlated noise, and  3) 
that it is able to nearly reach the poisson noise limit for our 30
minute AORs.  The heater cycling timescale for these datasets is of order 30 -
40 minutes.

\subsubsection{Snapshots of HD~158460}
\label{HD158460}
Our second line of evidence in favor of the snapshot technique comes
from a secondary dataset where we observed one of the IRAC calibration
stars, HD~158460, with a set of 18 snapshot observations.  HD~158460
was vetted by the IRAC project to be a non-variable source, and is
therefore used as part of the ongoing IRAC flux calibration effort
\citep{2005PASP..117..978R}.  It has a spectral type of A1Vn C.  The
goal of this experiment is to see if our snapshot technique could
recover the '€œtruth' light curve of this non-variable calibration star,
ie., a flat line as a function of time.  This test can't be performed
on any of the planet hosting stars because they have astrophysical
variability in their light curves (transits/eclipses/phase variations) whereas the IRAC calibrators have
been vetted by the IRAC team to not have variations
\citep{2005PASP..117..978R}.  Any variations from flat will indicate
inconsistencies in the snapshot technique.  These 18 snapshot
observations were observed and reduced in exactly the same way as
those of WASP-14b.

\begin{figure*}
\hspace*{-1cm}
\includegraphics[scale=0.5]{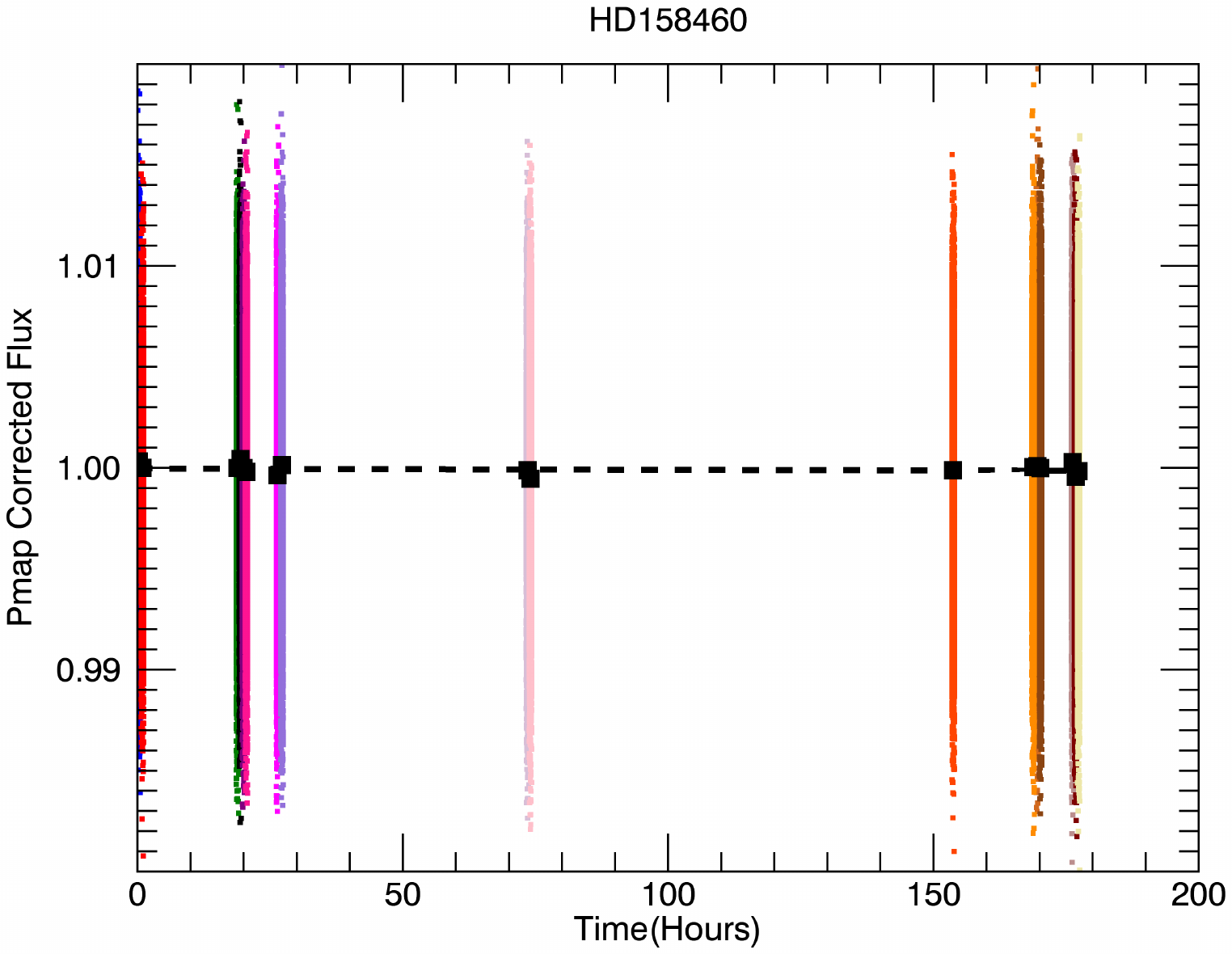}  
\hspace{-2cm}
\includegraphics[scale=0.5]{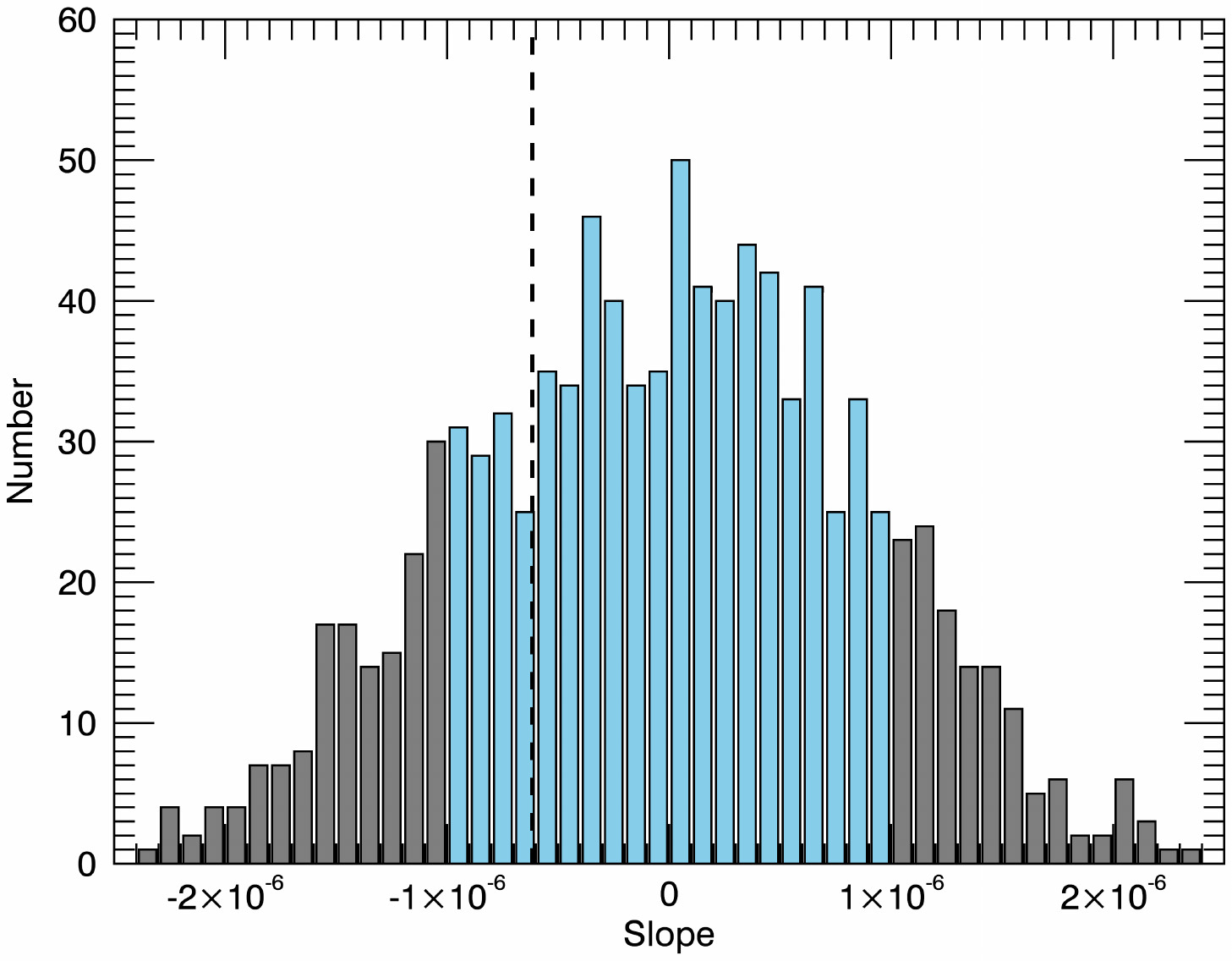}  
\caption[HD~158460]{ Snapshots of a calibration star.  Left: Light
  curve of a calibration star where each AOR is shown both unbinned
  with colored points and as a single binned black square.  Right: A monte carlo
  simulation of possible slopes of fitted lines to this data set.  The
  dashed line shows the chi-squared fit to the actual snapshot
  observations from the left plot.  FWHM of a gaussian fit to the
  histogram is indicated by a change in color from gray to blue.
  Snapshot observations are within one sigma of a flat light curve.
  See \S \ref{HD158460} for more details.}
\label{fig:HD158460}
\epsscale{1}
\end{figure*}

The left side of figure \ref{fig:HD158460} shows the corrected fluxes
of HD~158460 as a function of time over the $\sim200$ hours baseline of
observations.  Each colored data point represents an unbinned flux
measurement.  Black boxes show all the data from one AOR
binned together.  The dashed black line is the result of a chi-squared
linear fit to the binned datapoints.

To test the level to which this light curve is flat, we performed a
Monte Carlo simulation using 1000 instances of a Fischer-Yates shuffle
to randomly switch the time-stamps associated with each flux
measurement, and then re-measure the slope.  This shuffling must be done
on the binned data points, otherwise points from one AOR would get
shuffled into another AOR.  The right side of figure
\ref{fig:HD158460} shows a histogram of the slopes of resulting light
curves from the simulation.  The slope of the chi-squared fitted
linear fit to the snapshot dataset is shown with a dashed line.  The
histogram color changes from gray to blue inside of 1 FWHM.  The slope
of the measured fit to the snapshot dataset is within one sigma of
zero, demonstrating that the snapshot technique recovers a
light curve within one sigma of the 'truth' light curve,
which is a successful validation of the snapshot technique.

\subsection{Phase Curve Repeatability}
Because the snapshot dataset covers multiple orbits, any variation from one phase to the next
can be seen in the variation between snapshots that randomly landed
at similar phase in Figure \ref{fig:WASP-14b}.  Astrophysical
variation from phase to phase would imply we had observed exoplanet
weather across multiple periods.  We find that the maximum variation between
measurements (difference in average flux) at a single phase is $0.1\%$.  This implies that the
maximum variation from one phase to the next is less than or equal to
$0.1\%$.  The likely cause for the $0.1\%$ variation in the residuals
is the persistent images as discussed in section \ref{latent}, and
therefore we likely have detected no significant astrophysical
variation from one phase to the next.  To date there are no other
descriptions of measured phase to phase variations in the literature
as it is extremely difficult to disentangle the instrumental effects
from astrophysical effects at such low levels.  There is some
discussion in theoretical works of orbit to orbit variability,
predicted to be less than of order 1\% \citep{2010ApJ...720..344L,2013ApJ...767...76K,2015AREPS..43..509H}

\subsection{Transit and Secondary Eclipse Depth}
\label{eclipse}
\begin{figure*}
\hspace*{-0.5cm}
\includegraphics[scale=0.4]{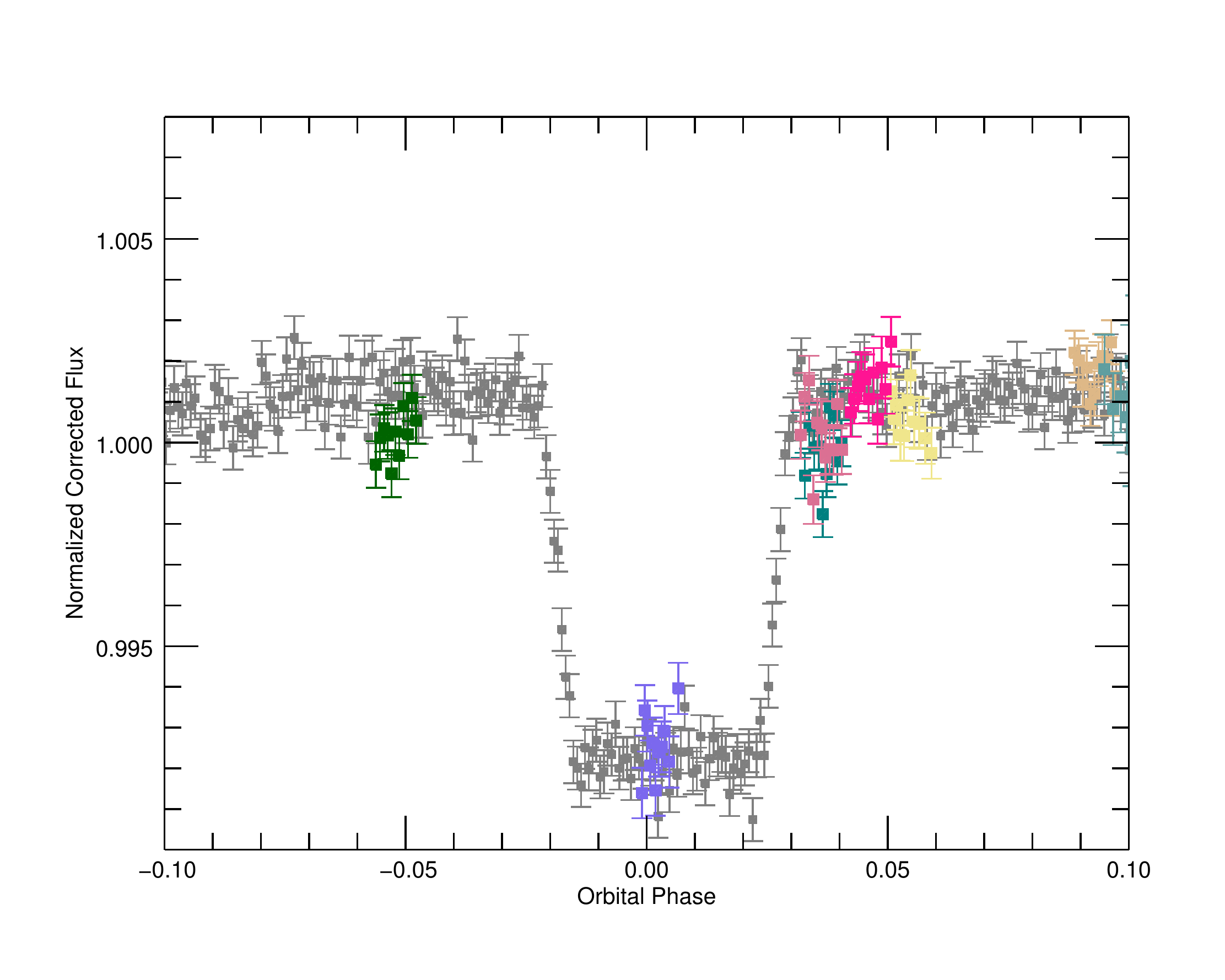}  
\hspace*{-0.5cm}
\includegraphics[scale=0.4]{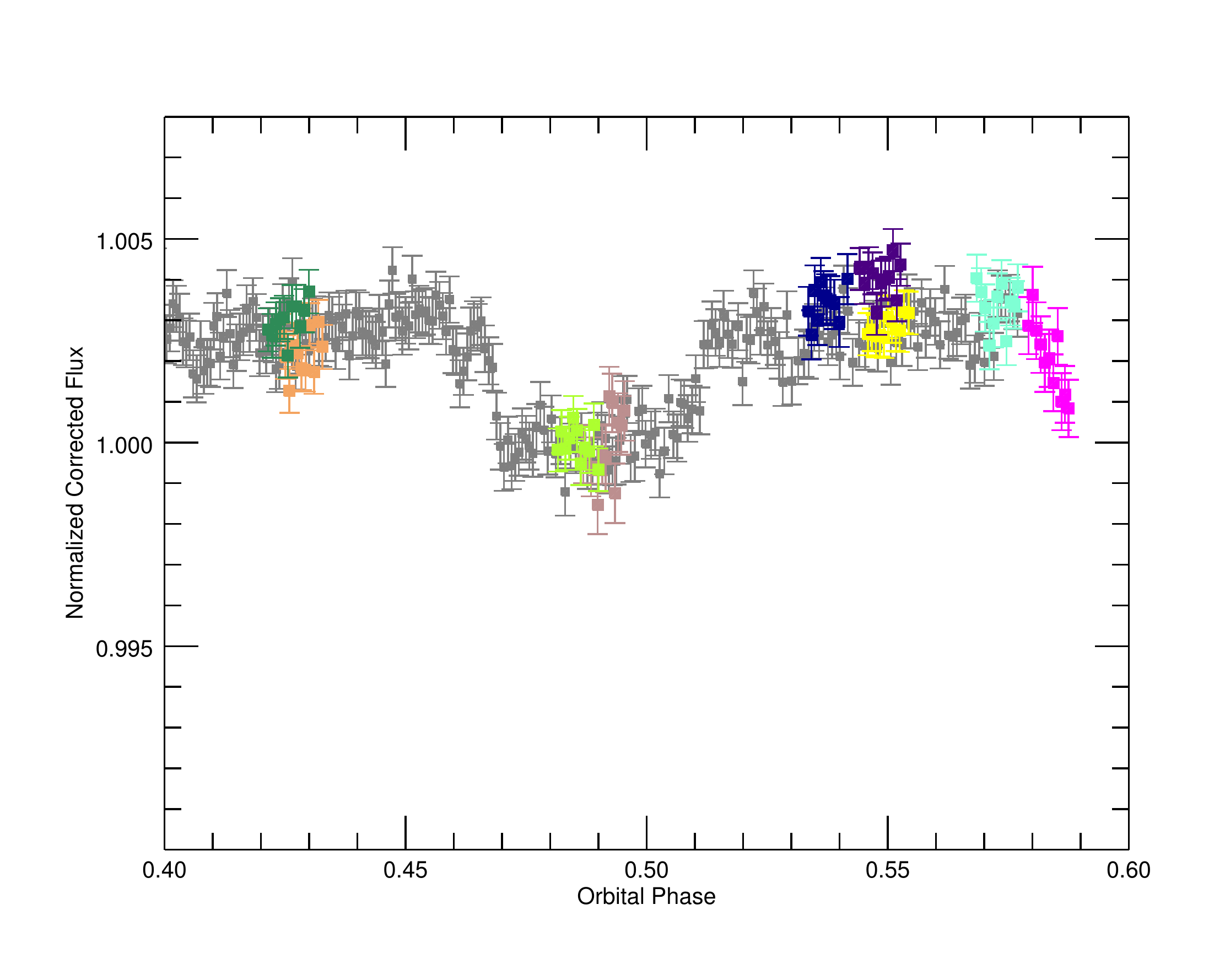}  
\caption[secondary]{Comparison of the the transit (left) and secondary
  eclipse (right) depths between snapshot and continuous staring mode data reduced
  in the same way.  Gray points are the continuous staring mode data, colored points are the snapshot
  data. Both plots have been normalized to the eclipse level. See \S
    \ref{eclipse} for more details.}
\label{fig:secondary}
\epsscale{1}
\end{figure*}

This project was not specifically designed to measure transit or
eclipse depths, however we do have snapshot data during transit and eclipse, so we
explore this topic briefly with the goal of seeing if we find eclipse
depth variation as a function of time in the 3.5 year baseline between
the first literature measurements and this paper.  Figure
\ref{fig:secondary} shows a zoom-in for comparison of our snapshot and
continuous staring mode data at both transit (left) and
eclipse(right). Our snapshot transit point is one of those
effected by the latent as noted in the superdark analysis, so while
our value is consistent with that in the literature, we make no
comment on the change in transit depth as a function of time.

We measure an eclipse depth (with two snapshots in eclipse) to be $0.222\pm0.07\%$.
The continuous staring mode data includes two secondary eclipses,
\citet{2015ApJ...811..122W} measure depths to be $0.2115^{+.0135}_{-
.0114}\%$ and $0.2367^{+ 0.0096}_{- 0.0142}\%$.  From a secondary eclipse
observation in March 2009 (3 - 3.5 years prior to the observations
presented here), \cite{2013ApJ...779....5B} measure the depth of the
secondary to be $0.224 \pm 0.018\%$. Our measured eclipse depth is
consistent with the range presented in those papers, and so we see no
evidence for an eclipse depth change as a function of time for
WASP-14b.


\section{Conclusion}
\label{conclusion}

In summary, we describe {\it Spitzer} IRAC snapshot high precision
photometry of WASP-14b.  We describe a new reduction technique which
is now provided for public use on the {\it Spitzer} Science Center website
in which we calibrate out the shape of the intrapixel gain using a
well-studied flat light curve star. We show three lines of evidence
that this technique is effective in measuring phase curves in a more
efficient manner than with continuous staring mode data.  Conclusions
specific to WASP-14b are that we confirm an eastward hot spot with an
amplitude of the phase curve between 0.00078 and 0.00098.  We put limits on the
phase to phase astrophysical variation of the system at less than $0.1\%$,
although that is probably an overestimate and due mainly to
instrumental systematics.  Lastly, we see no evidence for a change in
eclipse depth over a 3.5 year baseline from archival observations to
our most recent observations.

The largest limitation of this new technique is the effect of latent images
caused by previous observations on the stability of the
photometry.  For future snapshot observations we recommend observing a
dither pattern of a blank field after each snapshot to check for the
existence of persistent images and hopefully to remove that ``dark
pattern'' from the data.  To be even more safe, it may be advantageous
to observe all snapshots in the same visibility window so we are not
dealing with very different low level latents.  Of course, there can
always be local in time latents, but the first recommendation should
help with understanding those patterns.

It was useful to this paper to have a snapshot observation in transit
and in eclipse, mostly as checks on the phase curve shape.  For this
dataset it happened randomly but fortuitously.  In the future, it
might therefore be useful to specifically schedule an observation both
at transit and eclipse.


\acknowledgments

We thank the referee Nikole Lewis for useful suggestions which have
greatly improved the manuscript. This research has made use of data
from the Infrared Processing and Analysis Center/California Institute
of Technology, funded by the National Aeronautics and Space
Administration and the National Science Foundation.  This work was
based on observations obtained with the {\it Spitzer} Space Telescope,
which is operated by the Jet Propulsion Laboratory, California
Institute of Technology under a contract with NASA. This research has
made use of the NASA Exoplanet Archive, which is operated by the
California Institute of Technology, under contract with the National
Aeronautics and Space Administration under the Exoplanet Exploration
Program.  This research has made use of the NASA/ IPAC Infrared
Science Archive, which is operated by the Jet Propulsion Laboratory,
California Institute of Technology, under contract with the National
Aeronautics and Space Administration.  This research has made use of
the SIMBAD database, operated at CDS, Strasbourg, France.  This
research has made use of exoplanet.eu.  {\it Facilities:}
\facility{Spitzer (IRAC)}

\bibliography{ms_v8.bbl}  
\begin{deluxetable}{cccccccc}
\tablecolumns{7}
\tablewidth{0pc}
\tablecaption{Summary of Observations}
\tablehead{
\colhead{star} & 
\colhead{RA , DEC} &
\colhead{4.5 flux} &
\colhead{Dates} &
\colhead{Exptime} &
\colhead{\# of} &
\colhead{PIDs}\\
\colhead{} & 
\colhead{J2000} &
\colhead{mJy} &
\colhead{Observed} &
\colhead{(s)} &
\colhead{AORs} &
\colhead{}
}

\startdata
WASP-14b snaps   & 14:33:06.0 & 58 & 2012 Sep 07  &   2   & 46 & 80016 \\
                                &  +21:53:41            &    & 2013 Sep 16 &  &  & \\
WASP-14b stares   & 14:33:06.0 & 58 & 2012 Apr 24  &   2   & 5 & 80073 \\
                                &  +21:53:41            &    & 2012 Apr 26&  &  & \\
HD~158460 calstar   &17:25:41.3 & 965 & 2012 Jan 13 &   0.1   & 18 & 1331 \\
                                &  +60:05:23.8            &    &2012 Jan 20&  &  & \\
BD+67 1044 pmap   &17:58:54.7 & 642 & 2011 Mar &   0.1  & 332 & many\tablenotemark{a}\\
                                &  +67:47:36.9            &
                                &2012 Sep&  &  & \\

\enddata
\tablenotetext{a}{1320, 1326, 1328,1331, 1333, 1336, 1338, 1346, 1658, 1659, 1669}
\label{tab:obspars}
\end{deluxetable}

\begin{deluxetable}{ccccccccc}
\tablecolumns{3}
\tablewidth{0pc}
\tablecaption{Summary of Phase Curve Fitting Results}
\tablehead{
\colhead{Parameters} & 
\colhead{Wong et al. } &
\colhead{pmap-continuous} &
\colhead{pmap-snapshots} &
}

\startdata
Period(days)& 2.2437651& \nodata& \nodata\\
Inclination(degrees) &84.63 &\nodata &\nodata\\
$a/R_*$ &5.98 &\nodata &\nodata\\
k = $ecos(\omega)$ &-0.0247 &\nodata &\nodata\\
h = $esin(\omega)$ &-0.0792 &\nodata&\nodata\\
$T_0$&56042.687 &\nodata &\nodata\\
$R_p/R_*$&0.09421 &\nodata &\nodata\\
depth of 1 eclipse&0.002115&\nodata&\nodata\\
depth of 2 eclipse&0.002367&\nodata&\nodata\\
c1&8.06e-04$\pm$1.3e-05 &7.43e-04$\pm$1.0e-06 & 9.45e-04$\pm$2.9e-05\\
c2&1.30e-04$\pm$1.7e-05 & 2.17e-04$\pm$1.4e-06& 1.19e-04$\pm$3.9e-05\\
c3&5.23e-05$\pm$1.5e-05 &2.73e-05$\pm$1.2e-06 &-2.09e-04$\pm$3.7e-05\\
c4&4.51e-05$\pm$1.6e-05 &  8.97e-05$\pm$1.3e-06 & -1.21e-04$\pm$3.5e-05\\
ramp1&0&\nodata&\nodata\\
ramp2&0.07&\nodata&\nodata\\
\tableline
phase amplitude &7.86e-04$\pm$2.4E-5& 7.85e-04$\pm$1E-6 & 9.8e-04$\pm$2.85E-5\\
phase shift &9.4$\pm$2.5&16.15$\pm$2.5&11.0$\pm$2.5\\
\enddata
\tablecomments{Empty values in the table indicate those values are
fixed to the \citet{2015ApJ...811..122W} values.  We list
them here for completeness.}

\label{tab:fitpars}
\end{deluxetable}




\end{document}